\documentclass[twocolumn,tighten,usenames,dvipsnames,twocolappendix]{aastex631}
\usepackage[normalem]{ulem}
\usepackage{verbatim}
\usepackage{amssymb}
\usepackage{amsmath}
\usepackage{xspace}
\usepackage{graphicx}
\usepackage{lipsum}
\usepackage{multirow}

\usepackage{amsmath,mathtools,mathrsfs}
\usepackage{hyperref}
\usepackage{afterpage}
\usepackage{booktabs}

\newcommand{\msun}{{M_{\odot}}}
\newcommand{\erg}{{\rm erg}}
\newcommand{\s}{{\rm s}}
\newcommand{\yr}{{\rm yr}}
\newcommand{\K}{{\rm K}}
\def\kpc{{\rm kpc}}
\def\Mpc{{\rm Mpc}}
\def\km{{\rm km}}
\def\cm{{\rm cm}}

\newcommand{\VirA}{M87*\xspace}
\newcommand{\SgrA}{Sgr~A*\xspace}

\defcitealias{M87PaperI}{M87*~Paper~I}
\defcitealias{M87PaperII}{M87*~Paper~II}
\defcitealias{M87PaperIII}{M87*~Paper~III}
\defcitealias{M87PaperIV}{M87*~Paper~IV}
\defcitealias{M87PaperV}{M87*~Paper~V}
\defcitealias{M87PaperVI}{M87*~Paper~VI}
\defcitealias{M87PaperVII}{M87*~Paper~VII}
\defcitealias{M87PaperVIII}{M87*~Paper~VIII}
\defcitealias{M87PaperIX}{M87*~Paper~IX}
\defcitealias{SgrAPaperI}{Sgr~A*~Paper~I}
\defcitealias{SgrAPaperII}{Sgr~A*~Paper~II}
\defcitealias{SgrAPaperIII}{Sgr~A*~Paper~III}
\defcitealias{SgrAPaperIV}{Sgr~A*~Paper~IV}
\defcitealias{SgrAPaperV}{Sgr~A*~Paper~V}
\defcitealias{SgrAPaperVI}{Sgr~A*~Paper~VI}
\defcitealias{SgrAPaperVII}{Sgr~A*~Paper~VII}
\defcitealias{SgrAPaperVIII}{Sgr~A*~Paper~VIII}

\shorttitle{Excluding Naked Singularities in \SgrA and \VirA}
\shortauthors{A.E.~Broderick \& K.~Salehi}

\begin{document}

\title{Cosmic Censorship in Sgr A* and M87*: Observationally Excluding Naked Singularities}

\correspondingauthor{Avery E. Broderick}
\email{abroderick@perimeterinstitute.ca}

\author[0000-0002-3351-760X]{Avery E. Broderick}
\affiliation{Perimeter Institute for Theoretical Physics, 31 Caroline Street North, Waterloo, ON, N2L 2Y5, Canada}
\affiliation{Department of Physics and Astronomy, University of Waterloo, 200 University Avenue West, Waterloo, ON, N2L 3G1, Canada}
\affiliation{Waterloo Centre for Astrophysics, University of Waterloo, Waterloo, ON N2L 3G1 Canada}

\author[0009-0006-0070-1888]{Kiana Salehi}
\affiliation{Perimeter Institute for Theoretical Physics, 31 Caroline Street North, Waterloo, ON, N2L 2Y5, Canada}
\affiliation{Department of Physics and Astronomy, University of Waterloo, 200 University Avenue West, Waterloo, ON, N2L 3G1, Canada}
\affiliation{Waterloo Centre for Astrophysics, University of Waterloo, Waterloo, ON N2L 3G1 Canada}

\begin{abstract}
The imaging of Sagittarius A* (\SgrA) and the supermassive black hole at the center of Messier 87 (\VirA) by the Event Horizon Telescope constrains the location and nature of emission from these objects.  Coupled with flux limits from the near-infrared through the ultraviolet, the attendant size constraints provide strong evidence for the absence of an accretion-powered photosphere, and therefore for the existence of an event horizon about an astrophysical black hole.  Here, we demonstrate that a broad class of naked singularities exhibit inner turning points for time-like geodesics, and therefore may generically be excluded, regardless of the nature and unknown physical impact of singularity itself, subject to the single weak assumption that the its nongravitational impact is localized to its immediate vicinity.  While we restrict our attention to static, spherically symmetric spacetimes, we are nevertheless able to exclude or constrain a large number of commonly invoked naked singularity spacetimes in this way.
\end{abstract}


\section{Introduction}
\label{sec:intro}
Singularities are synonymous with black holes, the inexorable final destination of accreting material, and their resolution is among the chief motivations of the search for a quantum theory of gravity.  In the absence of such a theory, they remain a mystery, presenting an as yet unknown boundary condition for the spacetime and the fields that populate it, and a potential source of unknown physical effects.  For these reasons, predicting the future development of a singular spacetime and its contents presents a significant challenge.  Nevertheless, the possibility of creating naked singularities, i.e., singularities not hidden behind an event horizon, even if only in principle, by overspinning or overcharging general relativistic black holes remains unsettling.  Moreover, it remains unclear if it is possible to generate naked singularities from generic initial conditions in general relativity or alternate theories of gravity \citep[see, e.g.,][]{JNW:68,JMN-1:11,JMN-2:14}.

The formal problems associated with naked singularities threaten to render general relativity on its own a poorly posed initial value theory.  This unpalatable state of affairs may be ameliorated in the observable Universe if all singularities were to be hidden behind horizons (i.e., within black holes or behind the cosmic horizon).  This idea, codified as the ``cosmic censorship hypothesis'' introduced by \citet{Penrose:1969}, effectively ensures that no point within the Universe that we can currently observe is within the future Cauchy development of any singularity, and therefore unaffected by its unknown properties.  However, studies in higher-dimensional gravity provide strong evidence for generic violations of cosmic censorship \citep[see, e.g.,][]{Lehner:2010,Okawa:2011,Figueras:2017}, and example of generic violations have been identified in four-dimensional gravity \citep{JoshiDwivedi:93,JoshiMalafarina:2011,Cardoso:2018,Joshi:2019,Joshi:2020}.  Thus, it appears that the hypothesis in its original form is ruled out theoretically, rendering the problem only addressable experimentally.

Observational arguments for the existence of astrophysical event horizons\footnote{The teleological definition of an event horizon makes them impossible to study meaningfully by experiments on finite timescales; it would always remains possible that a signal could ``leak'' out moments after the experiment has concluded.  Nevertheless, by an ``astrophysical event horizon'' we mean a horizon (apparent or otherwise) that has existed for many dynamical times of the system, often much longer, and are therefore associated with many of the consequences of an event horizon for astronomical objects.  We will use this concept interchangeably with event horizon in what follows.} have existed for two decades now.  These typically invoke advection-dominated accretion flows, and the more general class of radiatively inefficient accretion flows \citep[RIAFs;][]{Narayan:1995,Narayan1998,BlandfordBegelman:1999,Yuan2014}.  Such flows are theoretically anticipated for accretion rates well below the Eddington rate, i.e., $\dot{M}<0.01 \,\dot{M}_{\rm Edd}=0.2\, (M/10^9~\msun)~\msun\,\yr^{-1}$, due to the weak Coulomb coupling between the electrons (which efficiently radiate) and the much more massive ions \citep[which liberate the vast majority of the gravitational binding energy;][]{Narayan1998,NarayanMcClintock:2008,Yuan2014}.  RIAFs, therefore, necessarily advect a large fraction of $\dot{M}c^2$ as kinetic energy toward the central object.  Wherein an event horizon (or at least apparent horizon) is present, this kinetic energy is lost, deposited within the black hole and increasing its mass.  However, in the absence of an event horizon, e.g., in the presence of a surface, this energy will typically be thermalized and radiated.  While thermalization happens naturally in baryonic atmospheres, for compact surfaces it is guaranteed by strong lensing for sufficiently high-redshift surfaces \citep{BLN09_sgra}.\footnote{See \citet{BN07_gravastars} for an example where the surface does not thermalize rapidly, and the potential for constraints on the existence of such systems even in that case.}  The result is a thermal bump in the spectral energy distribution (SED), the temperature and luminosity of which are set by energy balance with $\dot{M}$ \citep{NarayanGarciaMcClintock:1997,NarayanHeyl:2002,NarayanMcClintock:2008,BLN09_sgra,BNK15_m87}.

These thermal bumps are explicitly seen in the X-ray spectra of accreting neutron star X-ray binaries, there associated with emission from the boundary flow onto the stellar surface, and conspicuously absent in black hole X-ray binaries \citep{NarayanGarciaMcClintock:1997,NarayanHeyl:2002,McClintockNarayanRybicki:2004,NarayanMcClintock:2008}.  Supplemented with a long history of panchromatic observations, a detailed effort to understand and model accretion and jet launching in active galactic nuclei (AGN), the strongest constraints on visible surface/photosphere emission have been produced using the low-luminosity AGNs Sagittarius A* \citep[\SgrA;][]{BLN09_sgra} and Messier 87* \citep[\VirA;][]{BNK15_m87}.  Within these, the dominant systematic uncertainty was the size of the emitting surface; a large surface can be cooler and therefore escape detection by hiding underneath the bright emission at millimeter wavelengths from the much hotter accretion flow.

The Event Horizon Telescope (EHT) has effectively retired this uncertainty for \SgrA \citep[hereafter \citetalias{SgrAPaperI}-\citetalias{SgrAPaperVIII}]{SgrAPaperI,SgrAPaperII,SgrAPaperIII,SgrAPaperIV,SgrAPaperV,SgrAPaperVI,SgrAPaperVII,SgrAPaperVIII} and \VirA \citep[hereafter \citetalias{M87PaperI}-\citetalias{M87PaperIX}]{M87PaperI,M87PaperII,M87PaperIII,M87PaperIV,M87PaperV,M87PaperVI,M87PaperVII,M87PaperVIII,M87PaperIX} by directly imaging those sources on angular scales that resolve the putative event horizons.  These images confirm (1) that unstable circular photon orbits exist through the existence of the central brightness depression (see, e.g., \citealt{Broderick2023,Salehi2024}; \citealt{Kocherlakota:2021}; \citetalias{SgrAPaperVI} for implications beyond general relativistic black holes), and (2) that the emission region on the scale of the circular photon orbit is fully consistent with that anticipated by accretion onto general relativistic black holes \citepalias{M87PaperV,SgrAPaperV}.  Most importantly, the apparent size of any putative compact photosphere, in lieu of a horizon, is restricted to be smaller than the size of the observed shadow.  Thus it is EHT observations combined with contemporaneous near-infrared (NIR), optical and ultraviolet (UV) flux limits that provide the strongest empirical evidence currently for the existence of astrophysical event horizons.

The above empirical case in favor of event horizons is predicated on some degree of regularity near where the horizon would be, and may fail in the presence of a naked singularity.  For example, the unknown physical impact of the singularity itself on the accreting material may prevent the creation of a thermal photosphere, or accreting baryonic matter may simply disappear upon impacting the singularity.  Even should the accreted material remain for some time, the gravitational redshift may never be sufficiently high to support the argument that the emitting surface must approximately be in thermal equilibrium due to strong lensing.

Despite the inherent uncertainties, a number of authors have embarked on making theoretical predictions of the images from spacetime with naked curvature singularities \citep{JMN-1:11,VirbhadraEllis:2002,Joshi:2020,EichhornGoldHeld:2023,NguyenChristianChan:2023,EGB:24,Mishra:24,ChenWangPengYang:2024,Saurabh:2024}.  Typically, these images include a new family of lensed images, generated by ingoing photon trajectories that would otherwise be captured by the black hole.  On the basis of the excess interior flux such images would generate, EHT images of \SgrA have already been brought to bear on the existence of some naked singularity spacetimes \citepalias{SgrAPaperVI}.  Such limits presume that the intervening space between the singularity and the emission region is transparent, a potentially dubious assumption given the astrophysical properties of the EHT targets.  However, that generic, initially ingoing null geodesics escape the near-singularity region immediately raises the possibility of similar general behavior for time-like geodesics, and thus may impact the fate of the accreting baryonic material.

Given a single modest assumption about the physics of the singularity --- the direct unknown physical consequences of the singularity appear only within a very small distance of the singularity itself (e.g., the Planck length) --- we show that for a broad class of naked singularity spacetimes that the creation of a dense baryonic (and thus thermal) atmosphere is inevitable.  We describe how such an atmosphere develops for static, spherically symmetric, asymptotically flat spacetimes, and how its creation is related to the underlying gravitational properties of naked singularities.  For \SgrA and \VirA we further show that the existence of such an atmosphere is conclusively excluded, and therefore neither of these objects can harbor a naked singularity of a type in the previously mentioned class.

We begin in \autoref{sec:singularities} with a general discussion of static, spherically symmetric, asymptotically flat spacetimes exhibiting naked curvature singularities, presenting a classification scheme that identifies where and why the spacetime is singular.  In \autoref{sec:geodesics} we address the general behavior of null and time-like geodesics, and therefore the fate of accreting baryonic gas.  The spectral signatures and observational constraints for \SgrA and \VirA appear in \autoref{sec:signatures} and \autoref{sec:constraints}, respectively.  Finally, we collect concluding remarks in \autoref{sec:conclusions}.  Unless otherwise stated we assume a metric signature of $-+++$ and set $G=c=1$.

\section{General Stationary Spherically Symmetric Naked Singularities}
\label{sec:singularities}
For isolated accreting astrophysical black holes candidates, asymptotic flatness is well justified by observations of dynamics of test particles (e.g., stars and satellites) far from the object, i.e., in the Newtonian regime.\footnote{For objects with mass scales relevant for astronomical observations, we need not consider cosmological boundary conditions.}  For convenience, we will further restrict our attention to spherically symmetric spacetimes, leaving the treatment of rotating spacetimes for future work.  Within this class of spacetimes, staticity is a good approximation for systems where the mass accretion rate, $\dot{M}$, is sufficiently low that $M \dot{M}\ll M$.  For Eddington-limited accreting systems, $M \dot{M} \le 4\times10^{-3}\, (M/10^9 M_\odot)^2 M_\odot$, and thus this assumption is exceedingly well motivated for AGN and X-ray binaries (though possibly violated for gamma-ray bursts and certainly violated for black hole mergers).

The metric of any spherically symmetric, asymptotically flat, static spacetime can be written as
\begin{equation}
    ds^2 = -N^2 dt^2 + \frac{B^2}{N^2} dr^2 + r^2 d\Omega^2,
\label{eq:metric}
\end{equation}
where $r$ is the areal radius, and $N(r)$ and $B(r)$ are arbitrary functions of radius that asymptote to unity as $r\rightarrow\infty$. Well-behaved event horizons exist wherever $N(r)=0$; where $B(r)=\infty$ the coordinates becomes singular.  We turn now to the conditions under which \autoref{eq:metric} describes a spacetime with a naked singularity. 

A spacetime is singular if it contains incomplete time-like curves \citep[i.e., inextendible time-like trajectories,][]{Landsman:2021}, a notion closely related to geodesic incompleteness as described in \citet{Geroch:1968} and \citet{Penrose:1969}.  Such a circumstance may occur for a variety of reasons, including, for example, punctures or excisions of the spacetime manifold.  We focus exclusively on curvature singularities, like that found at the center of the Schwarzschild spacetime.  This has the virtue of being similar to and consistent with most of the literature on observational probes for singularities.  Henceforth, we will use ``singularity'' interchangeably with curvature singularity.

The Kretschmann scalar, $K\equiv R_{abcd} R^{abcd}$, for the spherically symmetric spacetime described by \autoref{eq:metric} is given by
\begin{multline}
    K = 
    \left\{
    \frac{1}{B}\left[\frac{\left(N^2\right)'}{B}\right]'
    \right\}^2
    +
    2 \left\{
    \frac{\left(N^2\right)'}{r B^2}
    \right\}^2\\
    +
    2 \left\{
    \frac{1}{r} \left(\frac{N^2}{B^2}\right)'
    \right\}^2
    +
    \left\{
    \frac{1}{r^2}\left(1-\frac{N^2}{B^2}\right)
    \right\}^2,
\label{eq:Kgen}
\end{multline}
\citep[see, e.g., eqs.\ 14-18 of][]{Gkig14}.  Singular behavior in $K$ necessarily betrays the presence of curvature singularities.  Because in spherical symmetry, $K$ is composed of the elements of the Riemann squared (the metric in \autoref{eq:metric} is manifestly diagonal), the converse statement is true: all spacetime curvature singularities will produce singular $K$.  When $B(r)=1$ everywhere (which is often true), \autoref{eq:Kgen} reduces to the more simple form
\begin{equation}
    K = 
    (f'')^2
    +
    \frac{4 (f')^2}{r^2} 
    +
    \frac{4 f^2}{r^4},
\label{eq:KN}
\end{equation}
where $f\equiv1-N^2$.  For Schwarzschild, $f=2M/r$, and the above reduces to the well-known result, $K=48 M^2/r^6$, which is singular at $r=0$. 

Therefore, the conditions upon the free functions in \autoref{eq:metric} for the spacetime to contain a naked singularity are:
\begin{enumerate}
\item at some radius (or collection of radii), $r_*$, $K$ becomes singular, implying that a spacetime singularity exists; and
\item for the outermost singularity, at all $r> r_*$ $N(r)$ and $B(r)$ are finite positive-definite functions of radius.
\end{enumerate}

\clearpage
\newlength{\tvs}
\setlength{\tvs}{0.185cm}
\begin{longrotatetable}
\begin{deluxetable*}{ccccccc}
\tabletypesize{\scriptsize}
\tablecaption{Example Spacetimes with Naked Singularities
\label{tab:spacetimes}}
\tablehead{
\colhead{Name\tablenotemark{a}} & 
\colhead{Type} & 
\colhead{Param.\ Range} &
\colhead{$N^2(r)$} & 
\colhead{$B^2(r)$} & 
\colhead{Null/Weak} &
\colhead{\hspace{0.5in} Reference \hspace{0.5in}}
}
\startdata
\\[-0.25cm]
RN & P0z & $|Q|>M$ & $\displaystyle 1-\frac{2M}{r}+\frac{Q^2}{r^2}$ & $1$ & Y/Y
& \citet{RN:1916}
\\[\tvs]
SHRN & P0z & $Q^2+s>M^2$ & $\displaystyle 1-\frac{2M}{r}+\frac{Q^2+s}{r^2}$ & $1$ & Y/Y
& \citet{SHRN:2013}
\\[\tvs]
EMD-1 & P0r & $q>\sqrt{2}$ & $\displaystyle 1- \frac{\sqrt{4M^2r^2+q^4}-q^2}{r^2}$ & $\displaystyle \frac{4r^2}{4r^2+q^4}$ & Y/Y
& \citet{Kocherlakota:2020}
\\[\tvs]
E ae-1 & P0z & $c_{13}<0$ or $c_{13}>1$ & $\displaystyle 1- \frac{2M}{r}-\frac{27 M^4 c_{13}}{16(1-c_{13}) r^4}$ & $1$ & Y/Y
& \citet{Kocherlakota:2020}
\\[\tvs]
E ae-2 & P0z & $\displaystyle \frac{2c_{13}-c_{14}}{1-c_{13}}<-8$ & $\displaystyle 1- \frac{(2-c_{14})M}{r} - \frac{(2c_{13}-c_{14})(2-c_{14})^2M^2}{8(1-c_{13})r^2}$ & $1$ & Y/Y 
& \citet{Kocherlakota:2020}
\\[\tvs]
EEH & P0z & $\alpha\le0$, $q_m^2>M^2$ & $\displaystyle 1-\frac{2M}{r}+\frac{q_m^2}{r^2}-\alpha \frac{2q_m^4}{5r^6}$ & $1$ & Y/Y
& \citet{EEH:2001}
\\[\tvs]
\multirow{2}{*}{JNW\tablenotemark{b}} & P0r & $0<\nu<1/2$ &
\multirow{2}{*}{$\displaystyle \left(1-\rho_*/\rho\right)^{\nu}$} & \multirow{2}{*}{$\displaystyle \frac{\left(1-\rho_*/\rho\right)^{\nu+1}}{ \left[ 1- (\nu+1)\rho_*/2\rho\right]^2}$} & \multirow{2}{*}{Y/Y} & \multirow{2}{*}{\citet{JNW:68}}
\\
& P0j & $1/2\le\nu<1$ & & & & 
\\[\tvs]
\multirow{2}{*}{JMN-1} & P0r & $0< M < R_b/3$
& \multirow{2}{*}{$\displaystyle \left(1-\frac{2M}{R_b}\right)\left(\frac{r}{R_b}\right)^{2M/(R_b-2M)}$} & \multirow{2}{*}{$\displaystyle \left(\frac{r}{R_b}\right)^{2M/(R_b-2M)}$} & \multirow{2}{*}{Y/Y}
& \multirow{2}{*}{\citet{JMN-1:11}}\\
& P0j & $R_b/3\le M < 2R_b/5$ & & & &
\\[\tvs]
JMN-2 & P0r & $0\le\lambda<1$ and $R_b>0$ & $\displaystyle \left[\frac{(1+\lambda)^2}{4\lambda\sqrt{2-\lambda^2}}\left(\frac{r}{R_b}\right)^{1-\lambda} - \frac{(1-\lambda)^2}{4\lambda\sqrt{2-\lambda^2}}\left(\frac{r}{R_b}\right)^{1+\lambda} \right]^2$ & $\displaystyle (2-\lambda^2)N^2(r)$ & Y/Y
& \citet{JMN-2:14}
\\[\tvs]
BST\tablenotemark{c} & P0r & $\beta,R_b\ne0$ & $\displaystyle \frac{2\beta^2}{1+R_b/r}$ & $\displaystyle N^2(r)/\beta^2$ & 
$\displaystyle\beta^2\le\frac{12}{13}$/$\displaystyle\beta^2\le\frac{2}{3}$
& \citet{BST:17}
\\[\tvs]
BINS\tablenotemark{d} & P0z & $a<4\bar{Q}^2$ and $\displaystyle M<M_e(a,\bar{Q})$ & $\displaystyle 1-\frac{2M}{r}-\frac{2\bar{Q}^2}{3r^2+3\sqrt{r^4+a\bar{Q}^2}}+\frac{4\bar{Q}^2}{3r^2} \bar{F}\left(\frac{a\bar{Q}^2}{r^4}\right)$ & 1 & Y/Y
& \citet{BIns:24}
\\[\tvs]
CNS & P0z & $Q\ne0$ & $\displaystyle \left(1+\frac{M}{r}\right)^{-2}+\frac{Q^2}{r^2}$ & 1 & Y/Y
& \citet{CNS:2024}
\\[\tvs]
4D EGB & P0r & $\gamma>0$ & $\displaystyle 1+\frac{r^2}{2\gamma M^2}\left(1-\sqrt{1+\frac{8\gamma M^3}{r^3}}\right)$ & 1 & 
Y/Y
& \citet{EGB:24}
\\[\tvs]
4D EGBQ & S1 & $|Q|>0$ and $\gamma>1-Q^2/M^2$ & $\displaystyle 1+\frac{r^2}{2\gamma M^2}\left(1-\sqrt{1+\frac{8\gamma M^3}{r^3}-\frac{4\gamma M^2 Q^2}{r^4}}\right)$ & 1 & Y/Y
& \citet{4DEGB:2020}
\\[0.5cm]
\enddata
\tablenotetext{a}{Abbrv.: RN--Reissner-Nordst\"om;
SHRN--Scalar hairy Reissner-Nordst\"om;
EMD-1--Einstein-Maxwell dilaton;
E ae-1,2--Einstein aether;
EEH--Einstein-Euler-Heisenberg;
JNW--Janis-Newman-Winicour;
JMN-1,2--Joshi-Malafarina-Narayan;
BST--Bertrand space-time;
BINS--Born-Infeld Naked Singularity;
CNS--charged naked singularity;
4D EGB--4D Einstein-Gauss-Bonnet;
4D EGBQ--Charged 4D Einstein-Gauss-Bonnet.}
\tablenotetext{b}{The function $s$ is defined implicitly by $r=\rho(1-\rho_*/\rho)^{(1-\nu)/2}$, where $\rho_*\equiv 2M/\nu$, and $\rho\in[\rho_*,\infty)$ maps smoothly onto $r\in[0,\infty)$ (see \autoref{app:JNW}).  Note that within the literature, the Janis-Newman-Winicour metric is also called the Fisher metric \citep{Fisher:48}, the Wyman solution \citep{Wyman:81,Virbhadra:97}, and the Buchdahl spacetime \citep{Buchdahl:1959,Bhadra:2001}.}
\tablenotetext{c}{Note that for the Bertrand singular spacetime, when $\beta=1$ it is the second term in \autoref{eq:Kgen} that is the lowest order term that diverges.}
\tablenotetext{d}{Born-Infeld naked singularity.  For compactness, we define $\bar{F}(z)\equiv\mbox{}_2F_1(1/4,1/2,5/4;-z)$, where $\mbox{}_2F_1(\dots)$ is the Gaussian hypergeometric function, and $M_e(a,\bar{Q})\equiv[4\bar{Q}^2-a+8\bar{Q}^2 \bar{F}(16a\bar{Q}^2/(4\bar{Q}^2-a)^2]/6\sqrt{4\bar{Q}^2-a}$.}
\end{deluxetable*}
\end{longrotatetable}

\subsection{Classification of Naked Singularities in Spherical Symmetry}
The condition that the spacetime is singular requires one of the four terms in \autoref{eq:Kgen} diverge at some radius $r_*$.  We classify such behaviors by their location and the lowest order of the derivative of the term that diverges.  First, by location, there are two possibilities.
\begin{description}
\item[P] if $r_*=0$, the singularity is ``point-like.''  This the only point-like singularity in spherical symmetry.
\item[S] if $r_*>0$, the singularity is ``shell-like.''  All non-point-like singularities are shell-like in spherical symmetry.
\end{description}
Second, by the lowest order of the first term that diverges:\footnote{It is possible for a naked singularity to be type P0 but the first term that diverges is not the last term in \autoref{eq:Kgen} if $\lim_{r\rightarrow0}N^2/B^2=1$.  Nevertheless, supplemented with the condition that a singularity does, in fact, exist, this does not present a significant confusion.},
\begin{description}
\item[0] $N^2/r^2 B^2 \rightarrow \infty$ at $r_*$, 
\item[1] $N^2/r^2 B^2$ is finite but $(N^2/B^2)'/r \rightarrow \infty$ or $(N^2)'/rB^2\rightarrow\infty$ at $r_*$, and
\item[2] $N^2/r^2 B^2$, $(N^2/B^2)'/r$, and $(N^2)'/rB^2$ are finite but $[(N^2)'/B]'/B\rightarrow\infty$ at $r_*$.
\end{description}
Thus a spacetime for which $N^2/r^2 B^2 \rightarrow \infty$ at $r_*=0$ is classified as P0.  Note that if any higher-order derivative term diverges at the origin, then the zeroth must as well (see \autoref{app:singularity_class}), and thus P0 is the only kind of ``point-like'' singularity in spherically symmetric, static spacetimes.

We further subdivide the type-0 spacetimes based on what is responsible for the divergence of $N^2/r^2 B^2$.
\begin{description}
    \item[z] If $N\rightarrow\infty$ at $r_*$ faster than $r^{-1}$, and $B^{-1}$ remains finite, we define this to be a ``redshift''-type singularity because the redshfit ($1+z=N^{-1}$) goes to zero at $r_*$.
    \item[j] If $B\rightarrow0$ at $r_*$ while $r$ and $N$ remain finite, we define this to be a ``Jacobian''-type singularity because $\sqrt{-g}=B r^2\sin\theta$ goes to zero at $r_*$.
    \item[r] If $r\rightarrow0$ at $r_*$ while $B$ and $N$ remain finite, we define this to be a ``radial''-type singularity, which necessarily only occurs for P0-type singularities.
\end{description}

In the following, we will demonstrate many key ideas with the Reissner-Nordstr\"om spacetime, which has the virtue of having or not having an event horizon, depending on the electric charge, $Q$, for which
\begin{equation}
    N^2(r) = 1-\frac{2M}{r}+\frac{Q^2}{r^2},
\end{equation}
and $B(r)=1$.  The associated Kretschmann scalar is
\begin{equation}
    K = \frac{48M^2}{r^6} - \frac{96MQ^2}{r^7} + \frac{56Q^4}{r^8},
\end{equation}
which is singular only at $r_*=0$, and therefore is type P0.  Because $B=1$ and $N\approx Q^2/r^2$ near $r=0$, this is further classified as P0z.  When $Q>M$ there is no event horizon and this metric provides a useful example.  A number of other naked singularity spacetimes are collected in \autoref{tab:spacetimes}, where their type, $N^2$, $B^2$, and relevant parameter ranges may be found.  For a handful of cases, identifying the singularity type requires careful assessment of the behavior of the metric near $r_*$ and may change with spacetime parameters (e.g., the JNW or JMN-1 spacetimes which are type P0r for some values and P0j for others).  We present explicit discussions of these specific spacetimes, i.e., JNW, JMN-1, JMN-2, and BINS, in \autoref{app:classification_of_specifics}.

\subsection{Energy Conditions and Singular Spacetimes}
\label{sec:energy}
For alternative gravity theories, the energy content of the spacetime may not be relevant.  Nevertheless, within the context of general relativity (and theories that admit an equation of the form of the Einstein equation), the null and weak energy conditions can provide insight into which classes of singular spacetimes are relevant and which may be ``sick.''

The consideration of energy conditions is motivated by \citet{Hawking92}, who proves that closed time-like curves (CTCs) cannot be produced by classical fields in compact regions of nonsingular spacetimes without violating the weak energy condition and conjectures that a similar condition holds for localized violations for quantum fields (the ``chronology protection conjecture'').  In the intervening three decades, generic violations have been found of the various energy conditions and their analogs averaged over time-like or null geodesics for both quantum and classical fields \citep{Visser:2000}.  One need not look to exotic spacetimes filled with wormholes and time machines for examples; such violations appear even in the modern standard cosmology due to dark energy \citep[see, e.g.,][]{Santos:2007}.  Nevertheless, despite not being generic, energy conditions remain convenient due to the many spacetime features that may be proven once they are assumed to hold \citep[for an extensive, if brief, list see][]{Curiel:2014}.

While the argument in \citet{Hawking92} does not apply in spacetime regions containing singularities, if we imagine that the naked singularity spacetime is the late-time evolution from nonsingular initial conditions, then the weak energy condition is sufficient to ensure that CTCs do not appear prior to the singularity's formation.  Therefore, insofar as exotic matter (negative rest energy) and CTCs are unphysical, the null and weak energy conditions provide some guidance as to which spacetimes may be relevant (apart, of course, having a naked singularlity!).

The Ricci tensor for the metric in \autoref{eq:metric} is diagonal and given by
\begin{equation}
\begin{aligned}
R^t_t = R^r_r &= 
-\frac{1}{2r^2B}\left[ \frac{r^2(N^2)'}{B} \right]'\\
R^\theta_\theta = R^\phi_\phi &=
\frac{1}{r^2} - \frac{1}{r^2 B} \left( \frac{r N^2}{B} \right)',
\end{aligned}
\end{equation}
from which we have the Ricci scalar
\begin{equation}
    R = -\frac{1}{r^2 B} \left[ \frac{(r^2N^2)'}{B} \right]' + \frac{2}{r^2}.
\end{equation}
The associated Einstein tensor is diagonal and given by
\begin{equation}
\begin{aligned}
G^t_t = G^r_r 
&= \frac{1}{r^2B}\left( \frac{rN^2}{B} \right)'-\frac{1}{r^2}\\
G^\theta_\theta = G^\phi_\phi
&= \frac{1}{2r^2B} \left[ \frac{r^2(N^2)'}{B} \right]',
\end{aligned}
\end{equation}
from which we may immediately obtain the stress-energy tensor via $T^\mu_\nu = G^\mu_\nu / 8\pi$.

The weak and null energy conditions must be met by all observers within the relevant class.  Because the spacetimes under consideration are spherical symmetric, without loss of generality, we restrict our attention to the equatorial plane, within which, we may write the momentum of an arbitrary observer as $p_\mu = (-e,X_r,0,\ell)$, which must satisfy
\begin{equation}
    p^\mu p_\mu = -\frac{e^2}{N^2} + \frac{B^2}{N^2} X_r^2 + \frac{\ell^2}{r^2} = -\mu^2,
\end{equation}
where $\mu>0$ for time-like observers and $\mu=0$ for null observers.  The statement of the null and weak energy conditions, $p_\mu T^\mu_\nu p^\nu \ge 0$ for all $p^\mu$, becomes
\begin{multline}
    -\frac{\mu^2}{r^2}\left[ \frac{1}{B}\left( \frac{rN^2}{B} \right)'-1\right]\\
    + \frac{\ell^2}{r^4} \left\{
      \frac{1}{2B} \left[
        \frac{r^4}{B} \left(\frac{N^2}{r^2}\right)'  
      \right]'
      +
      1
    \right\}
    \ge 0.
\end{multline}
That is, for the weak energy condition we require the coefficients of both $\mu^2$ and $\ell^2$ to be nonnegative for all $r>r_*$, while for the null energy condition we require only the latter.

Again, it is instructive to consider the case when $B(r)=1$, for which we find
\begin{equation}
    8\pi p_\mu T^\mu_\nu p^\nu
    =
    \mu^2 \left( \frac{f}{r^2} + \frac{f'}{r} \right)
    +
    \frac{\ell^2}{r^2} \left( 
    \frac{f}{r^2} - \frac{f''}{2} 
    \right),
\end{equation}
where $f$ is defined as in \autoref{eq:KN}.  The weak energy condition requires, in this case, that $f/r^2+f'/r\ge0$ and $f/r^2-f''/2\ge0$ at all $r>r_*$.   The first implies that for spacetimes exhibiting S1-type naked singularities and that obey the weak energy condition, $f'$ must diverge positively; were $f'$ to diverge negatively, $f/r$ must diverge as quickly, and the singularity would be of type S0.  Similarly, the second implies that for spacetimes exhibiting S2-type naked singularities and that obey the weak and/or null energy condition, $f''$ must diverge negatively (or be S0).  

For our example spacetime, Reissner-Nordstr\"om, these conditions reduce to
\begin{equation}
    \frac{f}{r^2} + \frac{f'}{r} = \frac{Q^2}{r^4} \ge 0
    ~~\text{and}~~
    \frac{f}{r^2} - \frac{f''}{2} = \frac{2Q^2}{r^4} \ge 0,
\end{equation}
both of which are satisfied unconditionally.  Therefore, the Reissner-Nordstr\"om spacetime satisfies the weak energy condition (and thus the null energy condition).  In \autoref{tab:spacetimes}, we note which naked singularity spacetimes pass which energy condition.

\section{Null and Time-like Geodesics about Spherically Symmetric Naked Singularities}
\label{sec:geodesics}
By definition, for a spacetime to contain a naked singularity, time-like geodesics that connect the singularity to infinity, $i_+$, must exist.  This does not mean that all time-like geodesics must extend from negative time-like infinity, $i_-$, to $i_+$.  Nor does it require that time-like geodesics not become generally trapped by the singularity itself.  In practice, however, for all stationary, spherically symmetric spacetimes with naked singularities of type P0z, P0r, or S0z, neither of these problems appear.  Apart from a set of finely tuned cases, comprising a set of measure zero in $(e,\ell)$ (see below), generically null and time-like geodesics escape the near-singularity region in finite time.  We now turn to demonstrating and quantifying these facts.

\subsection{Effective Potentials and Inner Turning Points}
All of the spacetimes described by \autoref{eq:metric} admit two constants of motion associated with the time-like and azimuthal symmetries.  As done in \autoref{sec:energy}, these are typically expressed in terms of an energy, $p_t=e$, and angular momentum, $p_\phi=\ell$.  The spherical symmetry implies that all motion is planar, and without loss of generality we may restrict our attention to motion confined to the equatorial plane.  Therefore, the final constant of motion is the effective mass of the underlying particle.

For photons, which are massless, the null condition is gives
\begin{equation}
    -\frac{e^2}{N^2} + \frac{B^2}{N^2} {p^r}^2 + \frac{\ell^2}{r^2} = 0,
\end{equation}
from which we have
\begin{equation}
    {p^r}^2 = \frac{1}{B^2} \left[ 
    e^2 - \frac{N^2 \ell^2}{r^2}
    \right]
    =
    \frac{1}{B^2} \left[
    e^2 - V^n_{\rm eff} (r,\ell) 
    \right],
    \label{eq:eom_null}
\end{equation}
where $V^n_{\rm eff}(r,\ell) = N^2\ell^2/r^2$ is the effective potential for null particles.  Whether or not photons will encounter an inner turning point, and therefore extend from $\mathscr{I}_-$ to $\mathscr{I}_+$, depends on which class of singularity is present and the photon's angular momentum.

For all type-0z and type-0r singularities (i.e., P0r, P0z, and S0z),  $V^n_{\rm eff}(r,\ell)$ diverges at $r_*$ by definition, and therefore for any $e$ and $\ell\ne0$, by continuity and asymptotic flatness, there must exist some $r>r_*$ at which ${p^r}^2=0$, and hence photons generally encounter an inner turning point.  For other types of singularities, the equation of motion of photons at the singularity may remain regular.  We will focus our attention on the type P0r, P0z, and S0z singularities henceforth unless otherwise stated.

\begin{figure}
    \centering
    \includegraphics[width=\columnwidth]{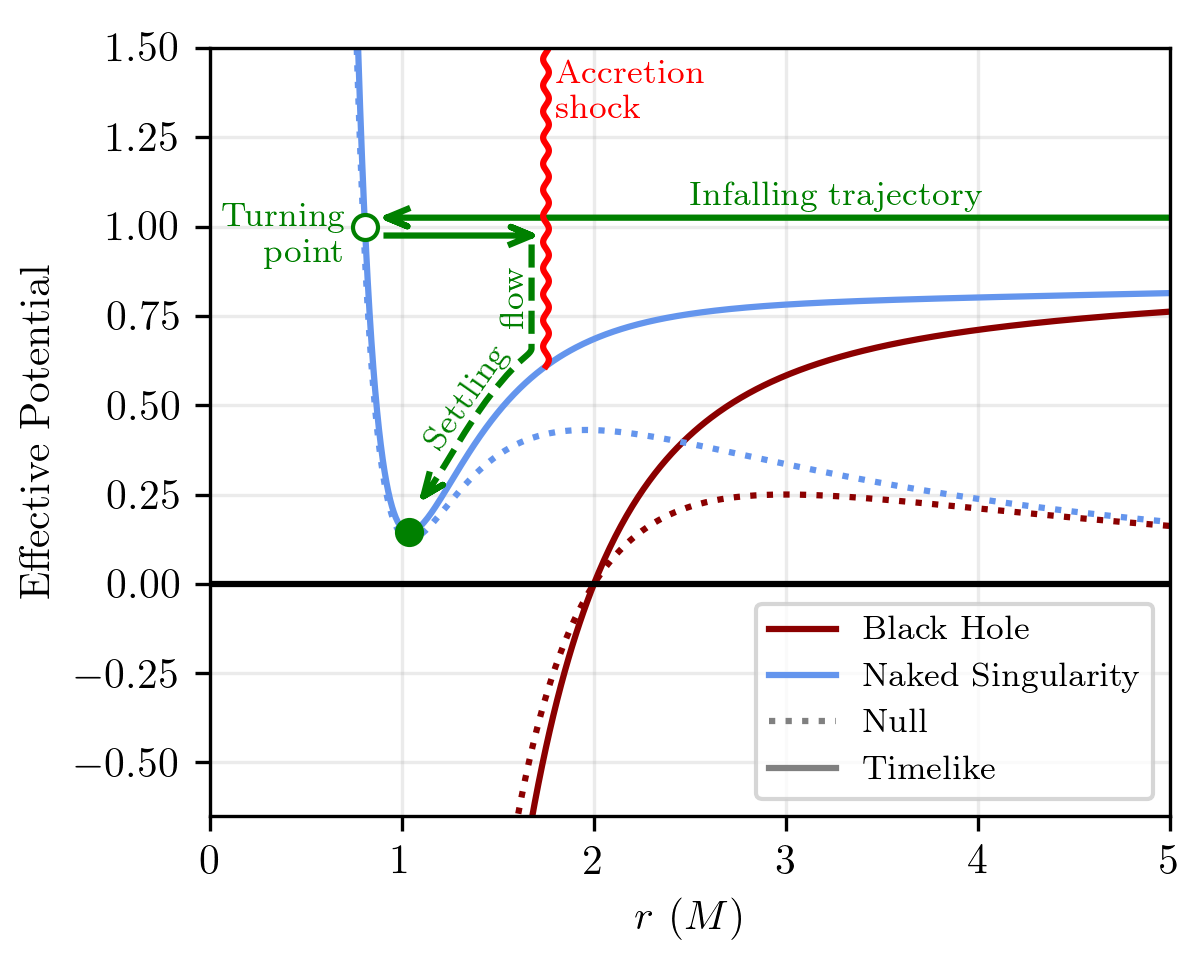}
    \caption{Null (dotted) and time-like (solid) effective potentials for Schwarzschild (dark red) and Reissner-Nordstr\"om (blue) spacetimes for $Q=1.01M$ and $\ell=0.5\sqrt{27}$. The trajectory for an infalling particle with $e=m$ at infinity is shown schematically in green, including the reflection at the inner turning point (open point), associated accretion shock when the outgoing and infalling flows collide, and the ultimate settling flow (dashed) depositing  material at the potential minimum (solid point).}
    \label{fig:veff_cartoon}
\end{figure}

For massive particles, the equation of motion departs from that of photons by the introduction of the particle mass, $m$ 
\begin{equation}
    {p^r}^2 
    =
    \frac{1}{B^2} \left[
    e^2 - V^t_{\rm eff} (r,\ell) 
    \right],
    \label{eq:eom_massive}
\end{equation}
where
\begin{equation}
    V^t_{\rm eff}(r,\ell) = V^n_{\rm eff}(r,\ell) + N^2 m^2
    \label{eq:Vt}
\end{equation} 
is the effective potential for massive particles.  Because $N^2$ is positive definite for all $r>r_*$ (because the singularity is naked) $V^t_{\rm eff}(r,\ell) > V^n_{\rm eff}(r,\ell)$ generally.  This has the significant consequence that anywhere photons encounter an inner turning point, massive particles with equal energy and angular momenta will do so as well at larger radii.  Massive particles with lesser energy will encounter an inner turning point even further away from the singularity (see, e.g., the examples in \autoref{fig:veff_cartoon}).  This fact has important consequences for the astronomical phenomenology of accreting systems.

\subsection{Time Delays at Inner Turning Points} \label{sec:dt}
Having established that for the nakedly singular spacetimes of interest (P0r, P0z, and S0z) that all time-like geodesics with $\ell\ne0$ will encounter a radial turning point prior to the singularity, we now turn to the practical question of how long this process takes as measured by distant observers.  Generally, the equations of motion for an outwardly moving massive particle are
\begin{equation}
    p^t = \frac{e}{N^2}
    ~~\text{and}~~
    p^r = \sqrt{\frac{e^2-V^t_{\rm eff}(r,\ell)}{B^2}}. 
\end{equation}
from which we have the time to propagate from $r$ inward to $r_{\rm tp}$ and back is, by symmetry
\begin{equation}
    \Delta t 
    =
    2 \int_{r_{\rm tp}}^r dr \frac{p^t}{p^r}
    =
    2 \int_{r_{\rm tp}}^r dr \frac{e B}{N^2\sqrt{e^2-V^t_{\rm eff}(r,\ell)}}.
    \label{eq:dt}
\end{equation}
For a black hole, the smallest value of $N^2(r)$ sets the timescale by virtue of the gravitational redshift.  Even in that case, the propagation time diverges only logarithmically with the maximum redshift (see \autoref{app:horizon_dt}).  For naked singularities, $N^2(r)$ never vanishes outside of $r_*$, and thus it is the radical in the denominator, which vanishes at the turning point, that sets the propagation timescale for massive particles.  

The integral in \autoref{eq:dt} may be evaluated approximately by Taylor expanding the radical term about the turning point.  That is, near $r_{\rm tp}$,
\begin{equation}
    e^2-V^t_{\rm eff}(r,\ell)
    =
    -{V^t_{\rm eff}}' (r-r_{\rm tp})
    +
    \dots,
\end{equation}
where ${V^t_{\rm eff}}'$ is the first derivative with respect to $r$ of $V^t_{\rm eff}(r,\ell)$ evaluated at $r_{\rm tp}$ and we have suppressed the dependence on $\ell$ for clarity and ignored the special case in which ${V^t_{\rm eff}}'$ vanishes.\footnote{For a given a particle energy, $e$, ${V^t_{\rm eff}}'$ will vanish at the turning point for orbits with angular momentum set by $\ell=2N\left(e^2-N^2m^2\right)^{3/2}/\left[ (N^2)' e^2 \right]$, where all quantities are evaluated at $r_{\rm tp}$.  For infinitesimally larger or smaller $e$ and/or $\ell$, ${V^t_{\rm eff}}'\ne0$.  Because we will be interested in the behavior of distributions of massive particles (i.e., accretion flows), we will ignore higher-order derivatives of $V^t_{\rm eff}$ in the estimation of propagation timescales as such terms are relevant for at most a set of measure zero in $(e,\ell)$.}  Therefore, the total time delay
\begin{equation}
    \Delta t 
    \lesssim
    \frac{e B(r_{\rm tp})}{N^2(r_{\rm tp})} \sqrt{\frac{r-r_{\rm tp}}{-{V^t_{\rm eff}}'}},
    \label{eq:dt_approx}
\end{equation}
is explicitly bounded with respect to $r_{\rm tp}$ (in stark contrast to the behavior near horizons, which typically diverges logarithmically with the distance from the horizon, see \autoref{app:horizon_dt}).  Thus, $\Delta t$ is generally finite.  

The magnitude of $\Delta t$ depends on the redshift at $r_{\rm tp}$, with larger redshifts corresponding to longer $\Delta t$.  However, in the absence of a naked singularity embedded in a very high-redshift region (i.e., one in which $N^2(r_{\rm tp})$ becomes very small, but never zero), the time delay is typically of order the light-crossing time of the system.  We quantify the size of the time delays required to appreciably modify the observational impacts of the return of accreting matter for specific sources in \autoref{sec:constraints}.

\section{Spectral Signatures of Naked Singularities}
\label{sec:signatures}
Summarizing the previous section: for P0r, P0z, and S0z singular spacetimes, infalling massive particles will generally encounter an inner turning point prior to reaching the singularity on a finite timescale as seen by a distant observer.  Importantly, this goes beyond proving the existence of a family of outward-going time-like geodesics (which exist by definition for naked singularities), but rather proves that generic infalling time-like geodesics will necessarily exit the near-singularity region, avoiding the singularity altogether.  This result has significant consequences for the spectral signatures of these spacetimes, set entirely by baryonic physics in the nonsingular portion of the spacetime.

We fashion an argument similar to \citet{BLN09_sgra} and \citet{BNK15_m87}, based upon the observed low efficiency of the accretion flows in \SgrA and \VirA, and constraints at shorter wavelengths on the thermalization of the unradiated kinetic energy.  This is predicated on four underlying assumptions and observations.

{\bf\em 1. The accretion flows on \SgrA and \VirA are radiatively inefficient.} 
That is, only a small fraction of the liberated gravitational binding energy is radiated during infall. For both sources this is both theoretically anticipated and observed. For objects with accretion rates below 0.1\% Eddington, Coulomb scattering is insufficient to redistribute energy from the ions, which by virtue of their mass accrue the majority of the gravitational binding energy, and the electrons, responsible for the overwhelming majority of the emission \citep{Narayan:1995,Narayan1998,NarayanMcClintock:2008,Yuan2014}. As a result, the vast majority of the kinetic energy of the accreting material is advected toward the central object.  In practice, the bolometric luminosity of \SgrA ($\simeq10^{36}~\erg\,\s^{-1}$) and \VirA ($\sim10^{42}~\erg\,\s^{-1}$) are well below estimates of $\dot{M}c^2$ ($\sim3\times10^{38}~\erg\,\s^{-1}$ and $10^{44}~\erg\,\s^{-1}$, respectively), arrived at by Faraday rotation limits and EHT observations.

{\bf\em 2. Interaction between inflowing and outflowing accretion streams efficiently thermalizes bulk flows.}  
After encountering the near-singularity turning point, accreting gas will generate a local, if bound, outflow.  For approximately axisymmetric accretion flows, which encompass all models under consideration for \SgrA and \VirA, these outflows will collide with subsequently accreting gas, though potentially at a distinct azimuthal location.  Once a counterflow is established (i.e., on an initial timescale of $\Delta t$ as defined in \autoref{eq:dt_approx}), the resulting, relativistic counterstreaming gas flows are subject to multiple mechanisms for thermalizing their bulk energy.  Chief among these mechanisms is an accretion shock that will thermalize the energy on the shock-crossing time, which is usually much less than the typical orbital timescale. 
Importantly, these dissipation mechanisms are dependent solely on the baryonic physics of the accreting material, and independent of the details of the singular spacetime.  Thus, it is generally anticipated that the kinetic energy of the bulk flow fully thermalizes on very short timescales.

{\bf\em 3. The thermalized region remains compact.}
Following thermalization, we will assume that the hot, thermalized region remains compact, i.e., smaller than the spacetime's photon orbit (should one exist).  This condition is a natural consequence of the very short thermalization timescale at the shocks, following which the gas will form into a settling flow toward the minimum of $V^t_{\rm eff}(r,0)$, $r_{\rm tp,min}$ which is generally outside $r_*$.  Moreover, in the cases of \SgrA and \VirA, the shadows observed by EHT indicate that the accretion flow remains well ordered down to the photon orbit, implying that any accretion shock must appear inside this region in those sources.

{\bf\em 4. The nongravitational impact of the singularity is localized to its immediate neighborhood.}
While the physics of the singularity is unknown, we presume that its nongravitational influence is felt only by material within some small characteristic length scale, $l_*$.\footnote{It may be natural to associate this scale with the Planck length, but it is not necessary.  For example, for the Reissner-Nordst\"om spacetime, $r_{\rm tp,min}\sim M$ for typical $\ell$ and when the singularity is naked.  This behavior is not confined to Reissner-Nordst\"om.  Therefore, $l_*$ may typically be much larger than the Planck scale and still be much smaller than $M$.}  While it may be possible to engineer spacetimes with naked singularities with $r_{\rm tp,min}-r_* < l_*$, for $l_*\ll M$, such situations represent an extreme fine tuning and we do not consider them further.

Note that we do not ignore nongravitational interactions between the accreting baryonic material and itself (indeed, these are a critical component in the second assumption listed above).  Furthermore, we do not make any assumptions about interactions with matter in the vicinity of the singularity, gravitational or otherwise.  Gravitational interactions, subject to spherical symmetry, are already incorporated into \autoref{eq:metric}.  Any forces due to nongravitational self-interactions (e.g., magnetohydrodynamic forces within the accretion flow itself) would need to diverge at the singularity to drive material into it, and can otherwise only shrink or expand the location of the turning point (only marginally if the acretting baryonic matter is nearly virialized). Similar physical restrictions apply to any physically viable matter distribution that is not part of the accreting gas.  While we do ignore the possibility that the accreting material vanishes at some finite distance from the singularity due to interactions with other matter, we do not view this as a significant limitation.

Subject to the above, the general result is an optically thick, compact settling flow that efficiently radiates at a rate set by that at which energy is advected inward.  That is, the luminosity and temperature as seen from a distant observer are
\begin{equation}
    L_\infty = 4\pi R^2 \sigma T_\infty^4
    ~~\text{and}~~
    T_\infty = \left( \frac{\eta_{\rm ad} \dot{M} c^2}{4\pi R^2 \sigma} \right)^{1/4},
    \label{eq:Tinf}
\end{equation}
where $\eta_{\rm ad}\gtrsim0.1$ is the fraction of the accreting rest mass energy that has liberated during infall that is advected into the settling flow, $R$ is the radius of the black hole shadow, and $\sigma$ is the Stefan-Boltzmann constant.

\section{Observational Constraints}
\label{sec:constraints}
The general development of an optically thick, baryonic photosphere around a broad class of naked singularities (P0z, P0r, and S0z) creates significant challenges for attempts to probe these spacetimes by looking for lensing signatures within the shadow \citepalias[e.g.,][]{SgrAPaperVI}.  Photons traveling on initially ingoing geodesics will be absorbed by the photosphere, which then prevents the additional lensing components from reaching distant observers.  Nevertheless, strong empirical constraints on singular spacetime can be obtained by looking for the thermal emission from these photospheres directly.

Because $T_\infty$ is set by $\dot{M}$, the associated thermal emission spectrum need not peak near 230~GHz, the frequency at which EHT observed \SgrA and \VirA in 2017 and 2018.  Indeed, for the inferred accretion rates in those sources, it peaks at much higher frequency, with the consequence that EHT observations alone place only a weak constraint on the existence of an additional shock-heated thermal component.  However, when the size constraints from EHT are supplemented with spectral flux measurements in the NIR, optical, and UV, such a component can be excluded altogether.  We now consider each target in turn, for which we summarize the state of the EHT observations and associated implied ranges for $T_\infty$, and compare to high-frequency flux measurements.

\subsection{\SgrA}
\begin{figure}
    \centering
    \includegraphics[width=\columnwidth]{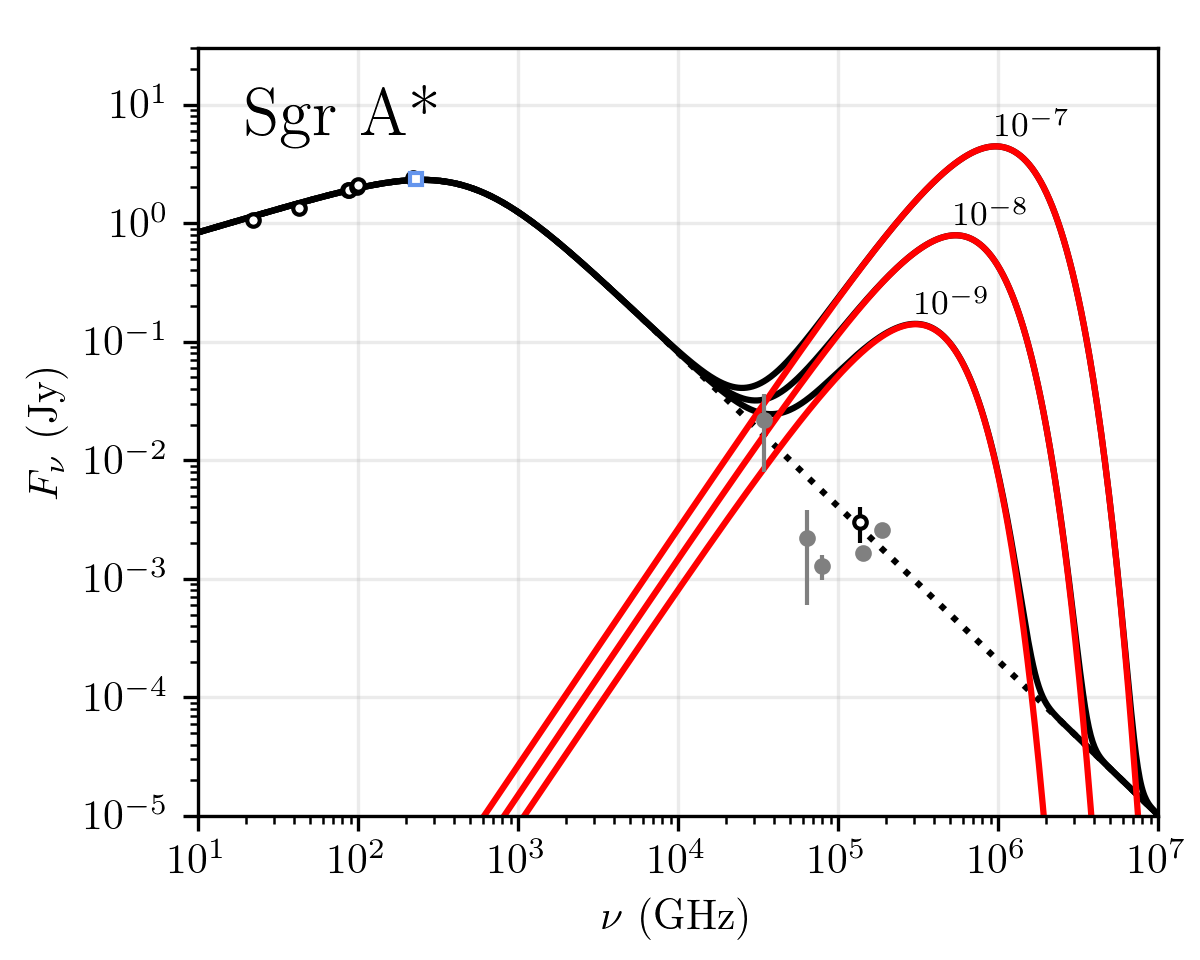}
    \caption{SED of \SgrA from a radially structured, nonthermal synchrotron-emitting region (black dotted) and a thermal bump due to an optically thick settling flow (red) for $\dot{M}$ ranging from $10^{-9}~\msun\,\yr^{-1}$ to $10^{-7}~\msun\,\yr^{-1}$.  For comparison, the 2017 EHT 230~GHz (1.3~mm) compact flux estimate \citep[open blue square;][]{SgrALightcurve:2022}, contemporaneous broadband SED \citepalias[open black points;][]{SgrAPaperII}, and historical NIR upper limits from \citet{BLN09_sgra} on the quiescent flux (gray closed points) are shown.}
    \label{fig:sgra}
\end{figure}
\SgrA is the bright radio source associated with the putative supermassive black hole at the center of the Milky Way.  Its SED, shown in \autoref{fig:sgra}, is well described by a RIAF that emits via synchrotron. Below $1\,$THz, \SgrA exhibits an inverted SED, with a spectral index ($F_\nu\propto\nu^{-\alpha}$) of $\alpha=-0.4$, characteristic of a self-absorbed, radially structured synchrotron source (\citealt{BlandfordKonigl:1979,BroderickLoeb2006b,Broderick2009b,Yuan2004}; \citetalias{SgrAPaperII}).  From NIR through the X-ray wavelengths, \SgrA's SED is well described by a power law with $\alpha=1.25$, typical of optically thin nonthermal synchrotron sources \citepalias{SgrAPaperII}. 

The mass of and distance to \SgrA are the best known of any black hole candidate, established most accurately via the orbits of individual stars that pass within $\sim10^{2-3}~{\rm au}$ of it, finding $M=(4.297+0.013)\times10^6~\msun$ and $D=8.277\pm34~\kpc$ \citep{GRAVITY:2022,Do:2019}.  However, that \SgrA is supermassive has been established in multiple ways.  It lies at the center of a stellar cusp, the structure and dynamics of which require a supermassive central object \citep{Genzel:2010}.  It sits still very nearly at the minimum of the Galactic potential, with a velocity toward the north Galactic pole of $-0.85\pm0.75~\km\,\s^{-1}$; if this is the result of dynamical friction it requires $M\gg1\times10^{6}~\msun$ \citep{ReidBrunthaler:2020}.  Similarly, the distance to \SgrA has been further verified by direct parallax measurements of masers in the Galactic center \citep{Reid:2019}.  This mass and distance is fully consistent with that inferred by EHT through directly imaging the 1.3~mm emission on Schwarzschild scales, which found $4.0^{+1.1}_{-0.6}\times10^6~\msun$ \citepalias{SgrAPaperI,SgrAPaperIV}.

The EHT images of \SgrA provide at least three lines of strong evidence for the canonical near-virial accretion flow model.  First, the presence of a dark shadow surrounded by a bright ring of emission matches quantitatively that anticipated by gravitational lensing about a $4\times10^6~\msun$ black hole, and in particular the existence of a circular photon orbit (\citetalias{SgrAPaperVI}; \citealt{Kocherlakota:2021,Broderick2023}).  Second, the shape and amplitude of the power spectrum of brightness fluctuations matches those associated with the turbulence responsible for angular momentum transfer through the accretion flow in by general relativistic magnetohydrodynamic (GRMHD) simulations (\citetalias{SgrAPaperV}; \citealt{Georgiev2022}).  Third, the linear polarization map is consistent with a predominantly toroidal magnetic field, again consistent with that anticipated by GRMHD simulations (\citetalias{SgrAPaperVII,SgrAPaperVIII}; \citealt{BroderickLoeb2006b}).  Therefore, while the EHT images of \SgrA do not preclude any interior emission,\footnote{The EHT is sensitive only to emission with brightness temperatures exceeding $\sim10^8~\K$, and thus would not be able to detect, e.g., the thermal emission from a settling flow surrounding a naked singularity \citep{BN06_sgra}.} they do confirm the absence of any obstruction beyond the photon orbit.

At $L=10^{36}~\erg\,\s^{-1}$, the total bolometric luminosity of \SgrA is well below its Eddington limit of $6\times10^{44}~\erg\,\s^{-1}$.  Despite this, the high observed Faraday rotation measure implies cold plasma densities at distances of $10$--$100M$ of $10^{6}~\cm^{-3}$, and corresponding to accretion rates of $\dot{M}\sim10^{-8}~\msun\,\yr^{-1}\approx10^{-7}\dot{M}_{\rm Edd}$, well into the RIAF regime \citep{Agol:2000,QuataertGruzinov:2000,Marrone:2007,Yuan2014}.  Similar conclusions follow from the direct modeling of the EHT images, which recover bolometric luminosities of $7-9\times10^{35}~\erg\,\s^{-1}$ with associated accretion rates of $0.5-1\times10^{-8}~\msun\,\yr^{-1}$ \citepalias{SgrAPaperV}.  Thus, the empirically derived radiative efficiency is at most $\eta\equiv L/\dot{M}c^2 \sim10^{-3}$, implying that the vast majority of the liberated gravitational potential energy is advected inward with the accretion flow.

\begin{deluxetable}{lcccccc}
\tabletypesize{\scriptsize}
\tablecaption{Time Delay Limits for \SgrA
\label{tab:sgra_delays}}
\tablehead{
\colhead{Observation/Phenomenon} & 
\colhead{$T$} & 
\colhead{$T/M$} & 
}
\startdata
Gyration period                 & 0.07 ms & $3\times10^{-6}$  \\
Coulomb relaxation time          & $\ll 0.6$~s & $\ll 3\times10^{-2}$ \\
\hline
Single EHT observation          & 8 hr  & $1\times10^3$ \\
EHT campaign duration           & 1 week  & $3\times10^4$ \\
Observational history of \SgrA  & 50 yr & $8\times10^7$ \\
Ionization echos                & 500 yr & $8\times10^{8}$ \\
Fermi bubbles                   & 1 Myr & $2\times10^{12}$ \\
Milky Way age                   & 10 Gyr & $2\times10^{16}$ \\
\enddata
\end{deluxetable}

After shock formation, the kinetic energy of the accreting matter is thermalized in two steps, both occurring on the shock-crossing timescale. First, the microscopic particle momenta isotropizes on the upstream gyration timescale
\begin{equation}
    t_G = 2\pi m_p c/eB \approx 0.7\, (B/{1~\rm G})^{-1}~{\rm ms}.  
\end{equation}
For a typical magnetic field of $10~{\rm G}$ in \SgrA, $t_G\sim3\times10^{-6} M$, and is therefore much less than the light-crossing time of the system.  Second, the baryon distribution function relaxes due to Coulomb scattering\footnote{It is possible for collective phenomena (e.g., Landau damping) to relax the baryon distribution even more rapidly.} on the timescale
\begin{equation}
\begin{aligned}
    t_{pp} &= \frac{\sqrt{2\pi}}{n \sigma_T c \ln\Lambda} \left(\frac{m_p}{m_e}\right)^2 
    \left(\frac{kT}{m_p c^2}\right)^{3/2} \\
    &\approx 0.6\, (n/10^6~{\rm cm}^{-3})^{-1} (T/10^4~{\rm K})^{3/2}~{\rm s},
\end{aligned}
\end{equation}
where $\sigma_T$ is the Thomson cross section, $\ln\Lambda\sim20$ is the Coulomb integral, $n$ in the baryon number density of the atmosphere, and $T$ is the atmosphere temperature \citep{MahadevanQuataert:1997}.  In this estimate, we have neglected the rapid growth in the baryon density associated with the collection of accreted material, and thus this is an upper limit on the relaxation timescale, which will be many orders of magnitude shorter in practice\footnote{The electron-baryon and electron-electron relaxation times are shorter by factors of $m_e/m_p$ and $(m_e/m_p)^2$, respectively.}  Thus, the limiting timescale is that associated with the development of a standing accretion shock, after which the kinetic energy of subsequently accreted material will thermalize nearly instantly.

A standing accretion shock will form on a timescale set by that over which accreted material initially reaches the inner turning point and returns outward to encounter the ensuing accretion flow.  This timescale is bounded, but nevertheless continues to be subject to the particulars of the metric through the coefficients in \autoref{eq:dt_approx} and could be larger than the light-crossing time (e.g., if the singularity is embedded within a region with very high redshift).  However, the relevant limit on this timescale for \SgrA depends on the accretion history of the source.  \autoref{tab:sgra_delays} lists a number of relevant timescales, over which we have varying degrees of evidence for accretion in \SgrA, typically with $\dot{M}$ far exceeding that inferred today.

The observed flux from \SgrA varies by $\sim50\%$ over a single EHT observation night (8~hr) and by a similar factor over the entire EHT campaign in 2017 (1~week), providing strong evidence for continuous accretion during these periods \citep{SgrALightcurve:2022}.  While substantially variable, \SgrA has been a bright point source since its identification 50~yr ago by \citet{BalickBrown:1974}, implying continuous accretion at similar $\dot{M}$ as those seen today.  The presence of Fe\textsc{i} K lines in the Galactic center are associated with X-ray echos, betraying a period of significantly higher activity $\sim500~{\rm yr}$ ago \citep{Koyama:2008,Nobukawa:2011,Ryu:2013,Churazov:2017,Chuard:2018}.  Evidence for activity at least over the past 1~Myr is found in the Wilkinson Microwave Anisotropy Probe haze \citep{Finkbeiner:2004} and associated Fermi Bubbles \citep{Dobler:2010,Su:2010}, which would suggest near-Eddington accretion rates at that time.  Age estimates of the stellar population in the Galactic disk extend to 10~Gyr \citep[see, e.g.,][]{Gallart:2024,Soderblom:2010}.  Assuming that \SgrA coevolved with the Milky Way, and during that time the black hole grew substantially via accretion, this sets a longest timescale of $\sim10$~Gyr.

The corresponding timescales in units of $M$ are also listed in \autoref{tab:sgra_delays} and are typically much larger than unity.  Thus, any spacetime that seeks to hide a naked singularity by delaying the thermalization of the settling flow must do so by many orders of magnitude, with at least a factor of $10^{12}$ to accommodate the well-established accretion history of \SgrA, and $10^{16}$ if we believe that \SgrA grew substantially throughout the cosmological history of the Milky Way.

From \autoref{eq:Tinf}, the temperature of an optically thick settling flow in \SgrA would be
\begin{equation}
    T_\infty \sim 10^4 \left(\frac{\eta_{\rm ad}}{0.1}\right)^{1/4} \left(\frac{\dot{M}}{10^{-8}~\msun\,\yr^{-1}}\right)^{1/4}~\K,
\end{equation}
which produces a thermal spectrum that peaks at optical wavelengths, shown for $\eta_{\rm ad}=0.1$ and various $\dot{M}$ in \autoref{fig:sgra}.\footnote{Variations in the central mass (and thus $R$) or the distance have negligible impact in comparison to the direct dependence on $\dot{M}$ for both \SgrA and \VirA.} For comparison, we also show the SED contemporaneous with the 2017 EHT observation campaign with a broken power-law fit approximating the self-absorbed synchrotron spectrum.  We supplement these with the historical limits on the quiescent NIR flux listed in Table 2 of \citet{BLN09_sgra}.  Optical flux estimates are not available due to the large extinction in the direction of the Galactic center.  Nevertheless, the NIR flux limits conclusively exclude the existence of such an additional thermal component.

\subsection{\VirA}
\begin{figure}
    \centering
    \includegraphics[width=\columnwidth]{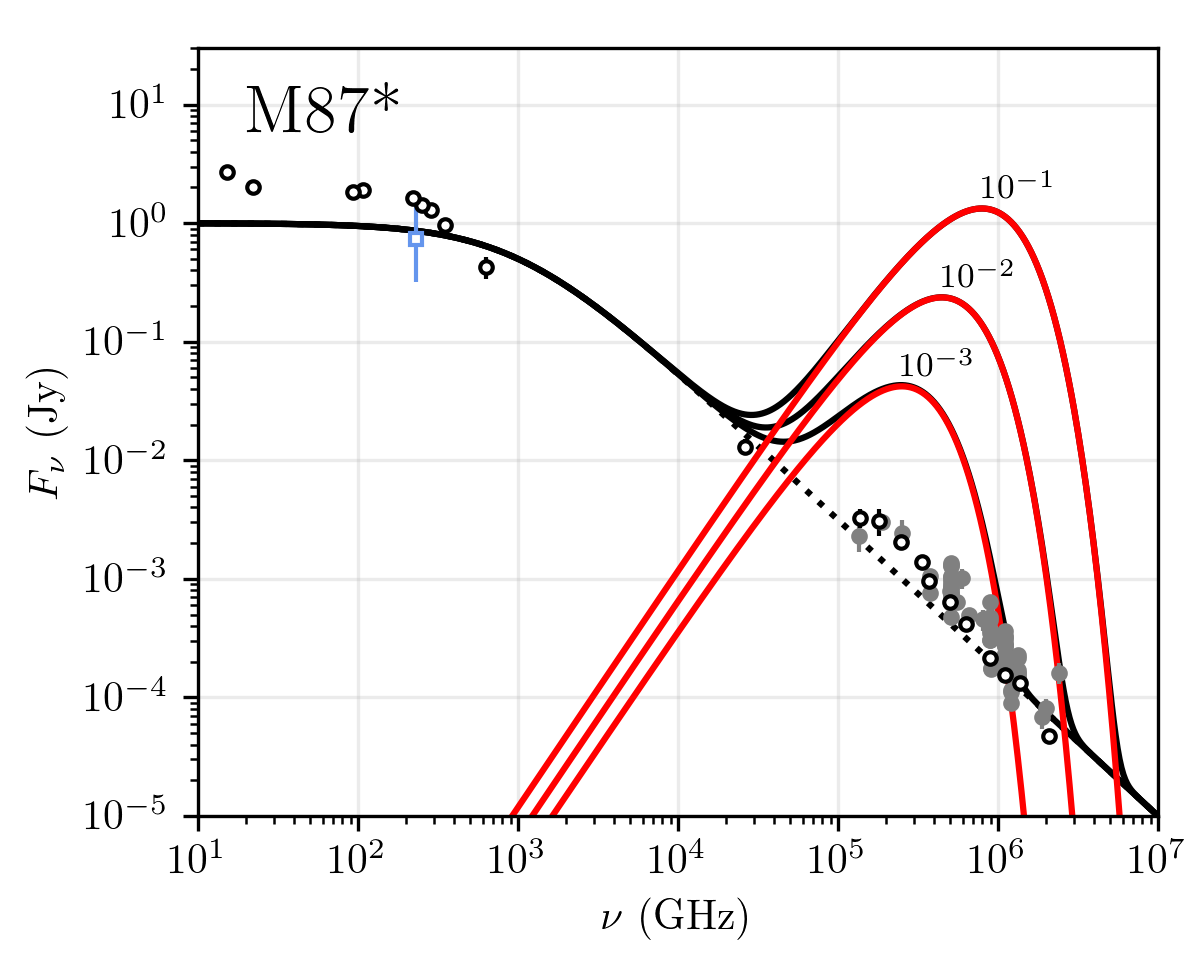}
    \caption{Spectral energy density of \VirA from a structured, nonthermal synchrotron-emitting region (black dotted) and a thermal bump due to an optically thick settling flow (red) for $\dot{M}$ ranging from $10^{-3}~\msun\,\yr^{-1}$ to $10^{-1}~\msun\,\yr^{-1}$.  For comparison, the 2017 EHT 230~GHz (1.3~mm) compact flux estimate \citepalias[open blue square;][]{M87PaperVI}, contemporaneous broadband SED \citep[open black points;][]{M87_MWL:2021}, and historical NIR-Optical-UV fluxes from \citet[][gray closed points]{BNK15_m87} are shown.}
    \label{fig:m87}
\end{figure}
The environment and behavior of \VirA differ from \SgrA in many relevant aspects.  It is located at the center of a giant elliptical galaxy, Messier 87, which is itself among the brightest galaxies in the Virgo cluster \citep{NGVCSV:2014}.  A bright, broadband radio source, \VirA sits at the bottom of a prominent jet, a narrowly collimated relativistic outflow which extends to $\sim30~\kpc$ \citep{deGasperin:2012}.  The origin of this jet is a topic of intense study, with both of the two main competing theoretical models powered by rotation and invoking twisted mangetic fields, either embedded within an orbiting accretion flow \citep{BP:1982} or in a black hole horizon \citep{BZ:1977}.\footnote{The Blandford-Znajek model invokes black hole spin, and thus violates the underlying assumption of spherical symmetry.  Nevertheless, we neglect spin further in our treatment here.}  Both models require a surrounding accretion flow to source and confine the magnetic field, and thus $\dot{M}$ is naturally related to the jet power.

Numerous independent estimates find a total jet power range of $\sim10^{44}$--$10^{45}~\erg\,\s^{-1}$ \citep[see Section 2.2 of][for a review]{BNK15_m87}. These estimates range across physical scale, and therefore temporal delay, from the $30~\kpc$ radio structure powered by the jet to HST-1, presumed to be powered by a jet shock \citep{deGasperin:2012,Stawarz:2006}.  This may be compared to $\dot{M}$ inferred from jet launching models, which under conservative assumptions are related by
\begin{equation}
    \dot{M} \gtrsim \frac{L_{\rm jet}}{2 c^2} \approx 10^{-3}~\msun\,\yr^{-1},
\end{equation}
\citep[see discussion in Section 2.1 of][]{BNK15_m87}.  This in close agreement with the $\dot{M}$ inferred from the Faraday rotation within the inner $40 M$, assuming that the Faraday screen lies within an RIAF \citep{Kuo:2014}.

The mass of \VirA has been measured primarily in two ways.  First, via the dynamics of the orbiting stellar population as inferred from their cumulative spectra.  This measurement is sensitive to assumptions regarding the dark matter cusp, and most recently found to be $6.14^{+1.07}_{-0.62}\times10^9~\msun$ \citep{Gebhardt2011}.  Second, through the dynamics of orbiting gas, which has typically found much smaller masses \citep{Walsh:2013}, though see \citet{Jeter:2019} and \citet{Jeter:2021} for a discussion of potential systematic uncertainties.  The larger, stellar dynamics mass is consistent with that inferred from the size of the bright ring observed by the EHT, $M=(6.5\pm0.7)\times10^9~\msun$, and the one that we make use of here \citepalias{M87PaperI,M87PaperVI}.  The distance is estimated either by using surface brightness fluctuations \citep{Bird:2010} or from the tip of the red giant branch \citep{Blakeslee:2009,cantiello2018next}.  We follow \citetalias{M87PaperVI} and combine these to arrive at the estimate $16.8^{+0.8}_{-0.7}~\Mpc$.

As with \SgrA, the EHT observations of \VirA are broadly consistent with the underlying astrophysical picture of the source.  The brightness asymmetry is in good agreement with the dynamics of the orbiting material and the orientation of the jet on larger scales.  The timescale over which EHT images of \VirA evolve is consistent with those found in GRMHD simulations, i.e., images  of  \VirA appear similar on neighboring days and differs on a week or longer (\citetalias{M87PaperIV,M87PaperV}; \citealt{M8718PaperI}).  The magnetic field geometry of the jet footprint in GRMHD simulations produces ``twisty'' linear polarization patterns that are quantitatively similar to that observed in EHT images \citepalias{M87PaperVII,M87PaperVIII}.  The size of the inner shadow is indicative of the strong gravitational lensing at the circular photon orbit, and like \SgrA, provides strong evidence that there is no interfering settling flow outside of this radius.

The SED of \VirA is well described by a self-absorbed synchrotron jet source that transitions from optically thick to thin near $1\,$THz, shown in \autoref{fig:m87} \citep{BlandfordKonigl:1979}. Below the transition frequency, the SED is essentially flat ($\alpha=0$); above the transition the SED is well fit by a power law with $\alpha=1.25$.  The isotropic bolometric luminosity of \VirA is approximately $10^{42}~\erg\,\s^{-1}$, with the bulk of that appearing in the submillimeter \citep{M87_MWL:2021}.  Were this completely due to accretion, for $\dot{M}=10^{-3}~\msun\,\yr^{-1}$, the radiative efficiency would be $\eta\sim10^{-2}$, reducing further if the jet is powered in part by some other source (e.g., spin of a naked singularity).  These estimates are in good agreement with the GRMHD modeling of \VirA based on the EHT images, which find $\dot{M}$ ranging from $3\times10^{-2}~\msun\,\yr^{-1}$ to $10^{-4}~\msun\,\yr^{-1}$ and an corresponding radiative efficiency range of $10^{-4}$--$10^{-1}$ \citepalias{M87PaperV}.  Thus, like in \SgrA, the majority of the liberated gravitational potential energy in \VirA is advected toward the black hole.  

\begin{deluxetable}{lcccccc}
\tabletypesize{\scriptsize}
\tablecaption{Time Delay Limits for \VirA
\label{tab:vira_delays}}
\tablehead{
\colhead{Observation/Phenomenon} & 
\colhead{$T$} & 
\colhead{$T/M$} & 
}
\startdata
Gyration period                  & 0.07 ms & $2\times10^{-9}$ \\ 
Coulomb relaxation time          & $\ll 5$~s & $\ll 2\times10^{-4}$ \\
\hline
Single EHT observation           & 8 hr   & $1$ \\
EHT campaign duration            & 1 week   & $2\times10^{1}$ \\
VLBI observations of \VirA       & 25 yr  & $2\times10^{4}$ \\
Observational history of M87 jet & 100 yr & $1\times10^{5}$ \\
Radio jet propagation time       & 200 kyr & $2\times10^{8}$ \\
Radio jet age                    & 40 Myr & $4\times10^{10}$ \\
M87 age                          & 10 Gyr  & $1\times10^{13}$ \\
\enddata
\end{deluxetable}

As with \SgrA, after the development of an accretion shock, in \VirA the accretion flow will thermalize on a shock-crossing time, set by the upstream gyration timescale.  In \VirA, the typical magnetic field strength is $10~{\rm G}$, and thus the gyration period is $t_G\sim 0.07~{\rm ms} = 2\times10^{-9} M$.  With a typical accretion flow baryon number density of $10^{-5}~{\rm cm}^{-3}$, the baryon relaxation time is also short, $t_{pp}\sim 5~{\rm s} \sim 2\times10^{-4} M$.  This estimate is a significant overestimate, falling sharply with the increasing atmosphere density resulting from accretion.  Therefore, the timescale over which an accretion shock will form is again set by the return time delay given in \autoref{eq:dt_approx}, which should be compared with the accretion history of \VirA.
 
Relevant timescales over which we have evidence for activity in \VirA are collected in \autoref{tab:vira_delays}.  Over the 2017 EHT campaign, the luminosity of \VirA was consistent with being constant, and thus over timescales of 8~hr to 1~week, the inferred $\dot{M}$ is constant.  In contrast to \SgrA, \VirA provides clear and continuous evidence for its activity via its jet.  Within $100~GM/c^2$, very long baseline interferometry (VLBI) observations over the past 25~yr have shown a bright core, confirming activity over this time \citep{Junor1999,Walker_2018}.  The jet has been observed on kiloparsec scales for over a century \citep{Curtis:1918}.  The jet itself contains a record of past activity, extending in each direction to 20~kpc in projection, which correspond to 70~kpc for an inclination of $17^\circ$ \citep{deGasperin:2012}.  Assuming the jets propagate relativistically, this implies an age from the jet propagation alone of more than 200~kyr.  The observed spectrum of the radio lobes into which the jets terminate implies the existence of a old population of energetic leptons with a synchrotron age of 40~Myr \citep{deGasperin:2012}.  Finally, globular clusters in M87 date back more than 10~Gyr.  \autoref{tab:vira_delays} lists the corresponding observational and astronomical timescales for \VirA.  In units of $M$, these timescales extend up to $10^{10}$ for those associated directly with the jet, and if \VirA coevolved with its host galaxy up to $10^{13}$.

For \VirA, \autoref{eq:Tinf} gives the equilibrium temperature for a putative optically thick settling flow to be
\begin{equation}
    T_\infty \sim 4\times10^3 \left(\frac{\eta_{\rm ad}}{0.1}\right)^{1/4} 
    \left(\frac{\dot{M}}{10^{-3}~\msun\,\yr^{-1}}\right)^{1/4}~\K.
\end{equation}
At this temperature, the associated thermal spectrum peaks in the NIR, though extends well into optical wavelengths.  This is shown for various $\dot{M}$ in \autoref{fig:m87}.  Measurements of the SED of \VirA made contemporaneously with the 2017 EHT campaign are also shown, along with a broken power-law fit approximating a self-absorbed synchrotron source.  We include the historical measurements from \citet{BNK15_m87} to provide some sense of the variation with time.

Because \VirA is not obscured at optical and UV wavelengths, the measured SEDs extend across nearly the entire range of frequencies relevant for the thermal peak.  Note that with the exception of the EHT compact flux measurement, all of the flux measurements are contaminated with emission from beyond $10M$ (e.g., HST-1).  Nevertheless, even with potential sizable contributions from outside the immediate vicinity of \VirA, the additional thermal component is conclusively excluded.

\section{Conclusions}
\label{sec:conclusions}
Naked singularities differ from other black hole foils in the potential for the introduction of unknown and uncontrolled physics, the effects of which must somehow be circumscribed.  Nevertheless, generally, a wide class of naked singularity spacetimes (P0z, P0r, and S0z) will contain inner turning points for time-like geodesics.  As a result, accreting baryonic gas will reverse direction in a short time as seen far from the object, rapidly encounter subsequently accreting gas, shock, and ultimately produce a settling flow outside of the singularity.  This process occurs entirely within the nonsingular part of the spacetime on scales that have already been observed; that it is independent of the unknown physics at the singularity corresponds to the modest assumption that the singularity's nongravitational impact is confined to matter that comes within some small distance of it.

That accreting material impinging upon the settling flow will thermalize the remainder of its kinetic energy, i.e., that gained during infall, is guaranteed by known baryonic physics.  Therefore, for this class of naked singularity spacetimes, an enveloping, optically thick, thermally emitting photosphere, is an inevitable consequence.  This photosphere may block initially ingoing null geodesics, rendering tests based on directly observing lensed emission inside the shadow challenging.  However,  for sources that are sufficiently radiatively inefficient and compact, this photosphere can be bright, and therefore its existence observationally probed.

\SgrA and \VirA satisfy both of these conditions.  Both sources are radiatively inefficient, i.e., only a small fraction of the liberated gravitational potential energy is radiated.  This is anticipated by theoretical models of the accretion process that have received significant empirical support from EHT observations and directly demonstrated via comparison of the source broadband luminosity to estimates of the accretion rate.  EHT observations of both sources also constrain the distance at which departures can occur from a black hole accretion flow to within the circular photon orbit.  That is, in both sources any putative baryonic photosphere would need to be compact.

The inferred $\dot{M}$ in both sources would generate an emitting photosphere with a temperature, as measured by a distant observer,  $\sim10^4~\K$, and therefore peaking in the NIR, optical, or UV.  The presence of such a thermal component can be conclusive excluded by flux measurements made contemporaneously with the EHT campaign in 2017.  That is, neither \SgrA nor \VirA can be a type-P0z, -P0r, or -S0z naked singularity spacetime.  Practically, this set of classes encompass all but one of the asymptotically flat, spherically symmetric, naked singularity spacetimes that we could find in the literature for at least some range of spacetime parameters (listed in \autoref{tab:spacetimes}).

There are at least three natural ways for naked singularities to evade the observational constraints we present.  First, the singularity may be of a different type, i.e., S1, S2, or P0j.  For example, the JNW and JMN-1 spacetimes are type P0j for a subset of spacetime parameters (see \autoref{tab:spacetimes}), and indeed have been found to not have generic turning points for time-like geodesics prior to reaching the singularity \citep{Patil:2012,JMN-1:11}.  Thus, for these spacetimes, we have merely narrowed the range of allowed parameters, placing a new upper limit on the strength of the scalar coupling in the JNW metric, corresponding to $\nu\ge1/2$, and a new upper limit on the transition radius in the JMN-1 metric, $R_b\le 3M$.

We note, however, that simply being a different type is not sufficient. While time-like geodesics in P0r, P0z, and S0z singular spacetimes have generic behaviors that result in now-excluded observational consequences, it is clearly possible for other types of naked singularity spacetimes to do so.  In \autoref{app:4DEGBQ}, we explicitly demonstrate that a type-S1 naked singularity spacetime (and the only example we could find in the literature) may be excluded by the argument described here.  Of course, we have also neglected any observable impact from the singularity itself.

Second, we have not considered the impact of spin.  Neither the class of naked singularity spacetimes we exclude nor the specific examples listed in \autoref{tab:spacetimes} are spinning.  Nevertheless, the formal possibility remains that the introduction of spin may make qualitative differences in the strength of constraints.  Incorporating spin complicates the question of the ultimate fate the accreting gas, substantially increasing of the size of the class of geodesics that must be surveyed.  Therefore, we leave an investigation into general classes of spinning spacetimes for future work (e.g., that presented by \citealt{Johannsen2016} and considered in \citealt{Salehi2024}).

Third, the unknown physics introduced by the singularity may include nonlocal interactions that extend past the inner turning point.  Were this to be the case, the chief advantage of the constraints we present, that they derive wholly from nonsingular regions within the spacetime, would be eliminated.  However, we consider this situation to be a rather different problem.

It remains possible for the time delay to be sufficiently high that a counterflow has not yet materialized. However, we view this as an extraordinary solution in that it requires the naked singularity be embedded in a region with sufficiently high redshift to increase the round trip time by more than a factor of $10^{10}$, effectively clothing the naked singularity and on astronomical timescales.

Finally, while we have argued that neither \SgrA nor \VirA can harbor a naked singularity of the proscribed types, this says nothing about whether or not naked singularities can exist in other sources.  Natural extensions of our argument can be made to other AGN and X-ray binaries in low-luminosity states (e.g., XTE J1118+480).  The main limitation will be, in the absence of images that resolve the gravitational radius, the need for an additional assumption regarding the compactness of the putative additional thermal component. Nevertheless, future ground-based \citep{ngEHT,ngEHT:2023} and space-based arrays promise to extend the reach of direct imaging to a much larger population of sources \citep{Pesce:2021,Pesce:2022}, and with it strong evidence excluding naked singularities in many additional sources.  

\begin{acknowledgments}
We thank Aaron Held for helpful discussions on the potential of indirect detection of naked singularities.  This work was supported in part by Perimeter Institute for Theoretical Physics.  Research at Perimeter Institute is supported by the Government of Canada through the Department of Innovation, Science and Economic Development Canada and by the Province of Ontario through the Ministry of Economic Development, Job Creation and Trade.  A.E.B. receives additional financial support from the Natural Sciences and Engineering Research Council of Canada through a Discovery Grant. 
\end{acknowledgments}

\bibliographystyle{aasjournal_aeb}
\bibliography{references,EHTCPapers}

\appendix

\section{Point-like Singularity Classes} \label{app:singularity_class}
For P-type singularities, necessarily the divergence of higher-order derivative terms imply that $N^2/r^2B^2$ diverges, and thus all P-type singularities are P0.  

We begin the proof of this by first looking at the case when $B=1$, for which the order class is set by the lowest-derivative term of $f/r^2$, $f'/r$, or $f''$ that diverges.  However, for a singularity to be anything other than P0 type, $f$ must vanish at $r=0$ sufficiently rapidly.  But in that case, by L'Hopital's rule
\begin{equation}
    \lim_{r\rightarrow0} \frac{f'}{r} 
    =
    2\lim_{r\rightarrow0} \frac{f}{r^2},
\end{equation}
and thus it cannot be type P1.  Similarly, if it is not P1,  $f'$ must vanish at $r=0$ but then
\begin{equation}
    \lim_{r\rightarrow0} f''
    =
    \lim_{r\rightarrow0} \frac{f'}{r},
\end{equation}
and it cannot be P2.  Thus, if a spacetime has a P-type singularity, has $B=1$, and is not type P0, then it cannot be P1 or P2, yielding a contradiction.  Because P-type singularities with $B=1$ exist, this immediately implies that all such singularities must be type P0.

We proceed with $B\ne1$ with similar line of argument.  A P-type singularity that is not type P0 requires that $N^2/B^2$ vanish at $r=0$ faster than $r^2$.  But in that case
\begin{equation}
    \lim_{r\rightarrow0} \frac{(N^2/B^2)'}{r}
    =
    \frac{1}{2} \lim_{r\rightarrow0} \frac{N^2/B^2}{r^2},
\end{equation}
and the first first-derivative term must also be finite.  If $B$ is nonzero at $r=0$, then, $N^2$ must vanish at $r=0$ and by L'Hopital's rule
\begin{equation}
    \lim_{r\rightarrow0} \frac{(N^2)'/r}{B^2}
    =
    \lim_{r\rightarrow0} \frac{N^2/r^2}{B^2}.
\end{equation}
If $B$ vanishes at $r=0$ then $N^2/r^2$ must also vanish at $r=0$ (which is a stronger statement than $N^2$ vanishing), and the above still applies.  Therefore, the second first-derivative term must also be finite.  That is, if a spacetime is not type P0, it cannot be type P1.

We repeat this argument to show that a spacetime that is neither P0 nor P1, it cannot be P2.  For the spacetime to not be type P1, $(N^2)'$ must vanish at $r=0$ at least as fast as $r$, and therefore, by L'Hopital's rule
\begin{equation}
    \lim_{r\rightarrow0} \frac{[(N^2)'/B]'}{B}
    =
    \lim_{r\rightarrow0} \frac{[(N^2)'/B]/r}{B},
\end{equation}
which immediately implies that it cannot be type P2. Finally, as before, because P-type singularities exist, this immediately implies that all such singularities must be type P0.

\section{Horizon Approach Timescale} 
\label{app:horizon_dt}
For comparison with the timescale estimates in \autoref{sec:dt}, here we make a similar estimate for general spherically symmetric, stationary black hole spacetimes.  Such spacetimes are also described by \autoref{eq:metric}, though defined by the vanishing of $N^2$ at some radius $r_h$.  Therefore, near this radius
\begin{equation}
    N^2(r) = (N^2)' (r-r_h) + \dots,
\end{equation}
where $(N^2)'$ is evaluated at $r_h$ (suppressed for clarity).  Thus, the timescale for reaching a radius $r=r_h+s$ for small $s$ is
\begin{equation}
\begin{aligned}
    \Delta t_h 
    &\approx \int_{r_h+s}^r dr \frac{eB}{(N^2)' (r-r_h) \sqrt{e^2-V^t_{\rm eff}}} \\
    &\approx \frac{eB}{(N^2)' \sqrt{e^2-V^t_{\rm eff}}} \ln\left(\frac{r-r_h}{s}\right).
\end{aligned}
\end{equation}
As noted in \autoref{sec:dt}, this diverges logarithmically with $s$. 

\section{Classification of Specific Singularities}
\label{app:classification_of_specifics}

For many of the spacetimes listed in \autoref{tab:spacetimes}, the classification of the singularity as described in \autoref{sec:singularities} is immediately clear from the expressions from $N^2(r)$ and $B^2(r)$.  However, for others, the classification process is more subtle, depending on the particular values of free parameters and/or requiring careful consideration of limits.  Here we explore these more challenging cases, taking each in turn.

\subsection{JNW}
\label{app:JNW}
The JNW metric as first presented by \citet{JNW:68} is not specified in areal coordinates, and thus not immediately expressible in the form of \autoref{eq:metric}.  Here we show how this may be done, at least implicitly, and identify the appropriate categorization of this metric.  We begin with the JNW metric as expressed in \citet{Kocherlakota:2020} and \citet{Patil:2012}
\begin{multline}
    ds^2 = -\left(1-\frac{\rho_*}{\rho}\right)^\nu dt^2 \\
    + \left(1-\frac{\rho_*}{\rho}\right)^{-\nu} d\rho^2
    + \rho^2 \left(1-\frac{\rho_*}{\rho}\right)^{1-\nu} d\Omega^2,
    \label{eq:JNW_metric}
\end{multline}
where $\rho_*<\rho<\infty$ is the radial coordinate and $0<\nu<1$ describes the strength of the coupling to the associated scalar field.  The quantity $\rho_*$ is uniquely set by $\nu$ and the Arnowitt–Deser–Misner (ADM) mass via $\rho_*=2M/\nu$.  This metric exhibits a curvature singularity at $\rho=\rho_*$.  The areal radius may be read off of the above, giving
\begin{equation*}
    r = \rho \left(1-\frac{\rho_*}{\rho}\right)^{(1-\nu)/2} = \rho^{\nu/2} \left(\rho-\rho_*\right)^{(1-\nu)/2}.
    \label{eq:JNW_r}
\end{equation*}
Both exponents are positive definite for the physically acceptable values of $\nu$ (i.e., for $\nu<1$).  This fact has two immediate consequences.  First $r$ vanishes when $\rho=\rho_*$, and thus the singularity occurs at $r=0$.  Second, both terms are smooth monotonically increasing functions of $\rho$, limiting to $\infty$ as $\rho\rightarrow\infty$.  That is, \autoref{eq:JNW_r} is a smooth, bijective map from $\rho\in[\rho_*,\infty)$ to $r\in[0,\infty)$, and therefore admits a unique and smooth inverse mapping from $r$ to $\rho$.  Even without explicitly constructing $\rho(r)$, we have
\begin{equation}
    d\rho = \left(1-\frac{\rho_*}{\rho}\right)^{(\nu+1)/2} \left[1-\frac{(\nu+1)}{2} \frac{\rho_*}{\rho}\right]^{-1} dr.
\end{equation}
Therefore, in terms of the areal radius, the JNW metric becomes
\begin{multline}
    ds^2 = -\left(1-\rho_*/\rho\right)^\nu dt^2 \\
    + \frac{\left(1-\rho_*/\rho\right)}{\left[1-(\nu+1)\rho_*/2\rho\right]^2} dr^2
    + r^2 d\Omega^2,
    \label{eq:JNW_metric2}
\end{multline}
as listed in \autoref{tab:spacetimes}.

It is now straightforward to show that $N^2/(B^2 r^2)$ diverges at $r=0$ (and $\rho=\rho_*$).  First, we note that near $r=0$, $1-\rho_*/\rho \approx (r/\rho_*)^{2/(1-\nu)}$.  Therefore
\begin{equation}
\begin{aligned}
    \lim_{r\rightarrow0} \frac{N^2}{r^2 B^2}
    &=
    \lim_{r\rightarrow0} \frac{\left[1-(\nu+1)\rho_*/2\rho\right]^2}{r^2 \left(1-\rho_*/\rho\right)} \\
    &=
    \lim_{r\rightarrow0} \frac{(1-\nu)^2\rho_*^{2/(1-\nu)}}{4 r^{(4-2\nu)/(1-\nu)}},
    \label{eq:JNW_K0}
\end{aligned}
\end{equation}
directly confirming that for $0<\nu<1$ the presence of a curvature singularity at $r=0$ of type P0.  The singularity subtype depends on which of the terms in the above is responsible for the divergence.  We note that
\begin{equation}
\begin{aligned}
    \lim_{r\rightarrow0} N^2 
    &=
    \lim_{r\rightarrow0} \left(1-\frac{\rho_*}{\rho}\right)^\nu \\
    &= 
    \lim_{r\rightarrow0} \left(\frac{r}{\rho_*}\right)^{2\nu/(1-\nu)},
\end{aligned}
\end{equation}
which vanishes for the full range of physically acceptable $\nu$, $0<\nu<1$.  However
\begin{equation}
\begin{aligned}
    \lim_{r\rightarrow0} \frac{N^2}{r^2} 
    &=
    \lim_{r\rightarrow0} \frac{1}{\rho_*^2} \left(\frac{r}{\rho_*}\right)^{(4\nu-2)/(1-\nu)},
\end{aligned}
\end{equation}
which diverges if $(4\nu-2)/(1-\nu)<0$.  Within the physically acceptable range of $\nu$, this occurs if $0<\nu<1/2$, for which the JNW singularity is of type P0r.  Alternatively, if $1/2\le\nu<1$, $\lim_{r\rightarrow0} N^2/r^2$ is bounded, and the JNW singularity is of type P0j.  This classification is consistent with the direct construction of geodesics in \citet{Patil:2012}, who find that finite angular momentum time-like geodesics can reach the singularity for $\nu\ge1/2$. 

\subsection{JMN-1}
The JMN-1 metric, first presented in \citet{JMN-1:11}, has a curvature singularity at $r=0$ of type P0 and is well posed for $0<M<2R_b/5$, where the upper limit arises from the condition that the sound speed within the collapsing material responsible for the singular spacetime is causal.  At $r<R_b$, $N^2(r)$ is a radial power law, with $N^2(r) \propto (r/R_b)^{\xi}$ (see \autoref{tab:spacetimes}).  The classification of the central singularity depends, therefore, on the power-law index, $\xi\equiv 2M/(R_b-2M)$, which itself is dependent upon the spacetime parameters $M$ and $R_b$.

Within the physically allowable range of $M$ and $R_b$, it is trivial to show that $\xi>0$, and thus the JMN-1 singularity cannot be of type P0z.  However, $N^2/r^2\propto(r/R_b)^{\xi-2}$ for $r<R_b$, which can diverge at $r=0$ if $\xi-2<0$, or equivalently $3M<R_b$.  Thus, for $0<M<R_b/3$, the JMN-1 singularity is of type P0r.  Conversely, for $R_b/3\le M<5R_b/2$, the singularity must be of type P0j.  We note that this also corresponds to the parameter range for which an unstable photon orbit is present \citep{Joshi:2020}.

\subsection{JMN-2}
The JMN-2 spacetime, first proposed by \citet{JMN-2:14}, contains a curvature singularity at $r=0$, as is clear from \autoref{tab:spacetimes}, and is thus of type P0.  The singularity subtype depends on the behavior of $N^2(r)$ near $r=0$.  For $0<\lambda<1$, at small radii, $N^2(r)$ is dominated by a term $\propto (r/R_b)^{2-2\lambda}$, and thus necessarily vanishes at $r=0$.  Thus, the spacetime cannot be type P0z.  However, because $\lambda>0$, $N^2/r^2$ is dominated by a term that diverges a $\propto (r/R_b)^{-2\lambda}$, and hence for these values of $\lambda$ the singularity is type P0r.  

The classification of the singularity as $\lambda$ approaches zero is more subtle, and requires a careful inspection of this limit of $N^2(r)$.  Taking the limit in $\lambda$ first, we find for $r<R_b$,
\begin{equation}
    \lim_{\lambda\rightarrow0} N^2 
    =
    \frac{1}{2\sqrt{2}}\frac{r}{R_b} \left[ 2 - \ln\left(\frac{r}{R_b}\right)\right].
\end{equation}
As $r$ becomes small, $\lim_{\lambda\rightarrow0} N^2$ vanishes, and thus the singularity cannot be type P0z.  However, at small radii $\lim_{\lambda\rightarrow0} N^2/r^2 \propto r^{-1}\ln(R_b/r)$, which diverges and thus again the singularity is type P0r.

\subsection{BINS}
The nature of the Born-Infeld naked singularity \citep{BIns:24} is dependent on the behavior of the hypergeometric function, $\bar{F}(z)$, in the limit of large argument, for which, the leading term is $\bar{F}(z)\approx\Gamma(1/4)\Gamma(5/4) z^{-1/4}/\sqrt{\pi}$.  Therefore, near $r=0$, the last term of $N^2(r)$ diverges as $\propto r^{-1}$ and thus the singularity is type P0z.

\section{Excluding 4D EGBQ}
\label{app:4DEGBQ}
Despite not being type P0, the 4D charged Einstein-Gauss-Bonnet (4D EGBQ) naked singularity spacetime is still excluded for all parameter values ($\gamma$, $M$, and $Q$), and provides a natural example of the application of the general argument made here to beyond the P0 class.

\begin{figure}
    \centering
    \includegraphics[width=\columnwidth]{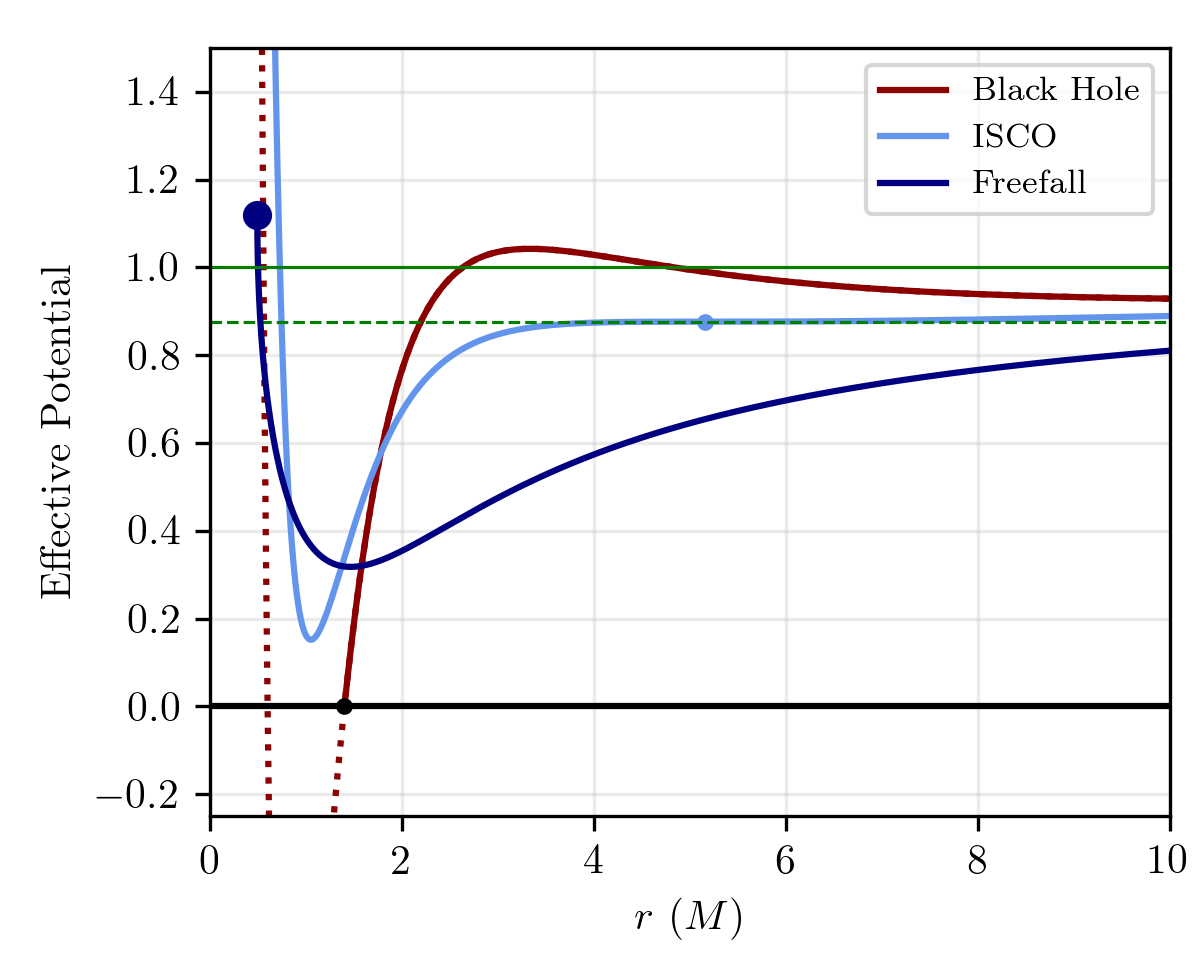}
    \caption{Effective potentials for three representative cases of the 4D EGBQ naked singularity with comparisons to the astrophysically relevant range of energies for accreting particles.  Shown are $V^t_{\rm eff}$ for orbits with $\ell=\sqrt{12}M$ in a black hole spacetime with $\gamma=0.8$ and $Q=0.2$ (dark red) with the radius of the horizon marked by the black point and the potential shown by a dotted line inside; orbits with $\ell$ given by that at the outermost innermost stable circular orbit (ISCO, light blue dot) in a naked singularity spacetime ($\gamma=1$ and $Q=0.2$, light blue); and orbits with $\ell=0$ in a naked singularity spacetime ($\gamma=1$ and $Q=1$, dark blue).  For comparison, in green the values of $e^2$ at the ISCO of the first singular spacetime (dashed) and $e^2=1$ (solid) are indicated.  While all examples have singularities at $r_*>0$, only in the last case is the singularity visible in the plotted range (large dark blue point).}    
    \label{fig:4DEGBQ}
\end{figure}

For the 4D EGBQ spacetime
\begin{equation}
    N^2(r) = 1 + \frac{r^2}{2\gamma M^2}\left(1-\sqrt{D}\right),
\end{equation}
where
\begin{equation}
    D \equiv 1 + \frac{8\gamma M^3}{r^3} - \frac{4\gamma M^2 Q^2}{r^4}.
\end{equation}
An event horizon exists when $\gamma\le1-Q^2/M^2$, and possesses a naked singularity otherwise.  The location of the singularity is set by the condition that $D=0$, and is the largest positive root of
\begin{equation}
    r_*^4 + 8\gamma M^2 r_* - 4\gamma M^2 Q^2 = 0.
\end{equation}
Because $\lim_{r\rightarrow\infty} D = 1$, by definition $D$ must be positive definite for all $r>r_*$, with two important consequences.

First, at $r=r_*$, $N^2(r_*) = 1 + r_*^2/(2\gamma M^2) > 1$.  A particle falling from infinity that is initially marginally bound has $e^2=m^2$, and thus even with $\ell=0$, $e^2-V^t_{\rm eff}(r_*,0)<0$ and there is an inner turning point for some $r>r_*$.  For accreting particles on orbits with $\ell\ne0$, $V^t_{\rm eff}(r_*,\ell)>V^t_{\rm eff}(r_*,0)$, and thus we concluded that an inner turning point always exists for physical accretion flows.  The effective potential for an example radial orbit is shown by the dark blue line in \autoref{fig:4DEGBQ}, along with the $e=1$ line, illustrating this behavior.

Second, $(N^2)'(r_*) < 0$, and hence $N^2(r_*)$ is a local maximum (in fact, it can be shown that $\lim_{r\rightarrow r_*} (N^2)'(r) = -\infty$.  Therefore, for accretion flows that effectively transport energy and angular momentum so that matter plunges from the outer ISCO, $e^2=V^t_{\rm eff}(r_{\rm ISCO},\ell_{\rm ISCO})<m^2$ and  $\ell_{\rm ISCO}\ne0$.  Hence, again, for such flows we generally have an inner turning point.  The effective potential for such a case is shown by the light blue line in \autoref{fig:4DEGBQ}, where $\ell=3.32M$ is the specific angular momentum at the ISCO, marked by the light blue dot.

Two practical effects may increase the specific energy of infalling orbits above unity.  First, the accreting matter may begin far from the black hole with nonzero kinetic energy, e.g., the Bondi radius is not at infinity.  However, such an increase is small -- neither \SgrA nor \VirA are accreting from a distant, relativistically hot plasma, and in both sources the brightness temperatures measured by the EHT, $\sim10^{10}~\K$, are less than the anticipated virial temperatures.  Moreover, even were this the case, and some accreting particles were sufficiently to surmount $V^t_{\rm eff}(r_*,0)$, a significant fraction of particles would necessarily be launched on orbits with $e<1$, and would therefore begin to form the inner settling flow, which when created would interact with and capture the energy from subsequently accreting particles of all masses.

\begin{figure}
    \centering
    \includegraphics[width=\columnwidth]{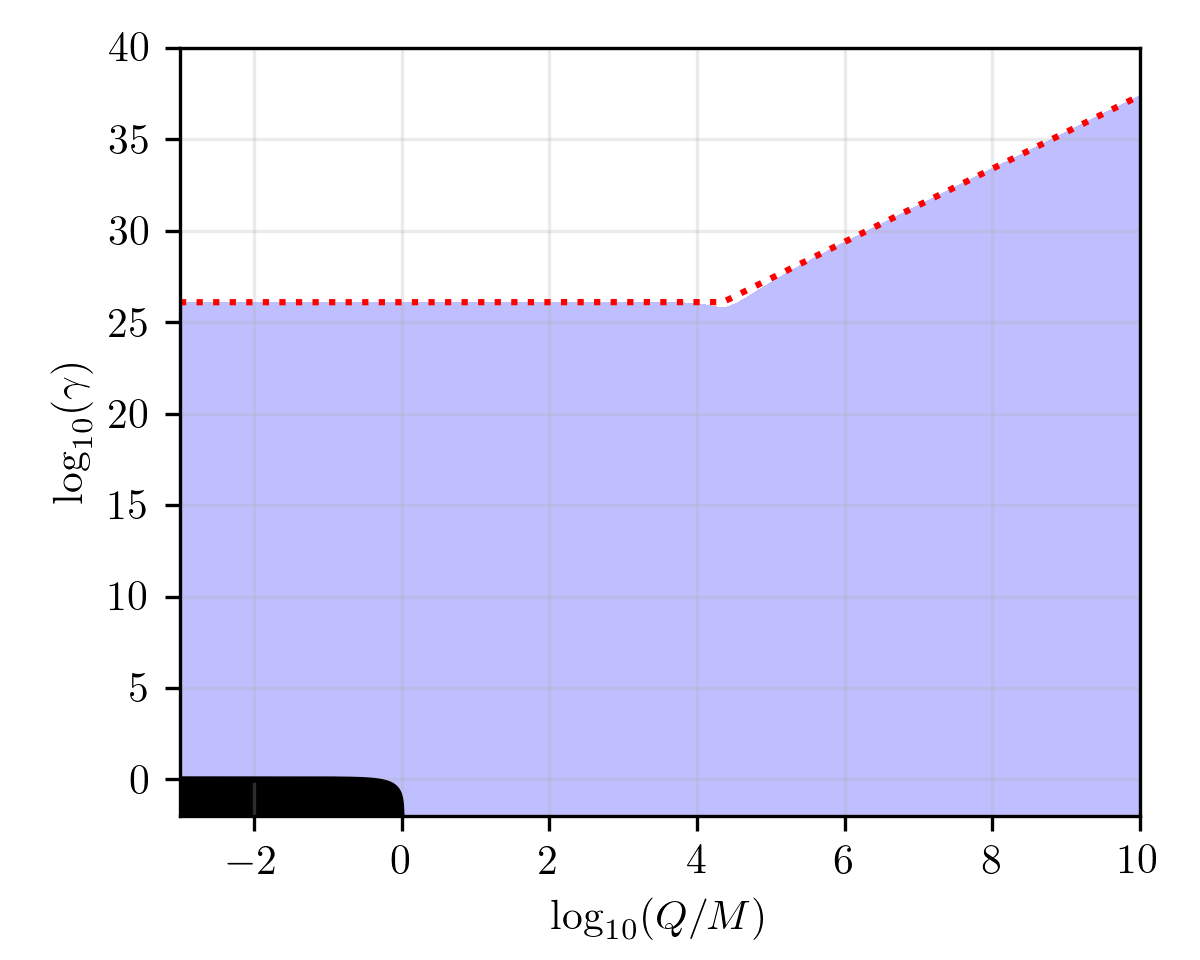}
    \caption{Excluded region (blue) in the $\gamma$--$Q$ parameter space of the 4D EGBQ naked singularity spacetime.  The black region in the lower left shows parameter values for which the 4D EGBQ metric describes a black hole.  The dotted red line shows the approximation in \autoref{eq:4DEGBQ_limits}.  Note that the axes are logarithmic.}
    \label{fig:4DEGBQ_limits}
\end{figure}

Second, the act of heating the settling flow via accretion necessarily increases $e$ locally.  Again, this effect is small in \SgrA and \VirA, neither of which are expected to have relativistically hot settling flows (i.e., $kT_\infty\ll m_p c^2$).  It remains formally possible for the thermal energy to be sufficient to raise the specific energy of the settling flow from the minimum to above $V^t_{\rm eff}(r_*,0)$. For small charges ($(Q/M)^6\ll64\gamma$), this condition is dominated by the minimum of the effective potential, which for large $\gamma$ occurs at $r\approx \gamma^{1/3} M$ and is $V^t_{\rm eff}(r,0)\approx1-\gamma^{-1/3}$.  For large charges ($(Q/M)^6\gg64\gamma$), it is dominated by the effective potential at the singularity, which occurs at $r_*\approx(4\gamma M^2Q^2)^{1/4}$ and $V^t_{\rm eff}(r_*,0)\approx \gamma^{-1/2} |Q/M|$.  Therefore, the energy gap to be overcome by the thermal energy is
\begin{equation}
    V^t_{\rm eff}(r_*,0) - V^t_{\rm eff}(r,0)
    =
    \begin{cases}
        \gamma^{-1/3} & (Q/M)^6\ll64\gamma\\
        \gamma^{-1/2} |Q/M| & (Q/M)^6\gg64\gamma,
    \end{cases}
\end{equation}
which implies that for the accretion heating of the settling flow to drive matter into the singularity, the coupling constant $\gamma$ must satisfy
\begin{equation}
    \gamma \gg 10^{26} \begin{cases}
        \displaystyle
        \left(\frac{T_{\infty}}{10^4~\K}\right)^{-3} 
        &
        \displaystyle
        \frac{Q}{M} < 2\times10^4 \left(\frac{T_{\infty}}{10^4~\K}\right)^{-1/2}\\
        \displaystyle
        \left(\frac{T_{\infty}}{10^4~\K}\right)^{-2} \left(\frac{Q}{M}\right)^{2} 
        & 
        \displaystyle
        \text{otherwise.}
    \end{cases}
    \label{eq:4DEGBQ_limits}
\end{equation}
This approximate constraint on $\gamma$ matches that obtained by direct numerical computation, shown in \autoref{fig:4DEGBQ_limits}, and effectively excludes the entirety of the physically credible parameter space.

\end{document}